\documentclass{sig-alternate}
\usepackage{graphicx}
\usepackage{balance}
\usepackage{times}
\usepackage{soul}
\usepackage{listings}
\usepackage{xspace}
\usepackage{subcaption}
\usepackage{csvsimple}
\usepackage{algorithm}
\usepackage[noend]{algpseudocode}
\usepackage{enumitem}
\usepackage{multirow}
\usepackage{tabu}
\usepackage{booktabs}
\usepackage[group-separator={,}]{siunitx}
\newcommand{\graphgen}{{\sc GraphGen}\xspace}
\newcommand{\topic}[1]{\vspace{-3.5pt}\smallskip \smallskip \noindent{\bf #1:}}
\newcommand{\topicul}[1]{\smallskip \noindent\underline{\bf #1:}}

\newcommand{\cutforrevision}[1]{\ignorespaces}

\newcommand{\calV}{\mathcal{V}}

\begin{document}
\title{Extracting and Analyzing Hidden Graphs from Relational Databases}


%
%
%

\numberofauthors{2} 

\author{
%
%
\alignauthor
Konstantinos Xirogiannopoulos\\
       \affaddr{University of Maryland, College Park}\\
       \email{kostasx@cs.umd.edu}
\alignauthor Amol Deshpande\\
\affaddr{University of Maryland, College Park}\\
       \email{amol@cs.umd.edu}
}


\date{30 July 1999}

\maketitle
\begin{abstract}

Analyzing interconnection structures among underlying entities or objects in a dataset through the use of graph analytics can provide
tremendous value in many application domains. However, graphs are not the primary representation choice for storing most data today, and in order to
have access to these analyses, users are forced to \textit{manually extract} data from their data stores, {\em construct} the requisite graphs, and
then {\em load} them into some graph engine in order to execute their graph analysis task. Moreover, in many cases (especially when the graphs are
        dense), these graphs can be \textit{significantly larger} than the initial input stored in the database, making it infeasible to construct or
analyze such graphs in memory.  In this paper we address both of these challenges by building a system that enables users to \textit{declaratively}
specify graph extraction tasks over a relational database schema and then execute graph algorithms on the extracted graphs.
We propose a declarative domain specific language for this purpose, and pair it up with a novel {\em condensed}, in-memory representation that
significantly reduces the memory footprint of these graphs, permitting analysis of larger-than-memory graphs. We present a general algorithm
for creating such a condensed representation for a large class of graph extraction queries against arbitrary schemas.
We observe that the condensed representation suffers from a {\em duplication} issue, that results in inaccuracies for most graph algorithms.
We then present a suite of in-memory representations that handle this duplication in different ways and allow \textit{trading off} the memory
required and the computational cost for executing different graph algorithms. We also introduce several novel \textit{deduplication}
algorithms for removing this duplication in the graph, which are of independent interest for graph compression, and provide a comprehensive
experimental evaluation over several real-world and synthetic datasets illustrating these trade-offs.\footnote{A shorter version of this paper appeared in ACM SIGMOD 2017.}

\end{abstract}

\section{Introduction}
\label{sec:intro}

Analyzing the interconnection structure, i.e., {\em graph structure}, among the underlying entities or objects in a dataset can
provide significant insights and value in many application domains such as social media, finance, health,
and sciences. This has led to an increasing interest in executing a wide variety of graph
analysis tasks and graph algorithms, e.g., community detection, influence propagation, network evolution, anomaly
detection, centrality analysis, etc. Many specialized graph databases (e.g.,
Neo4j, Titan, OrientDB, etc.), and graph execution engines (e.g., Giraph, GraphLab, Ligra, Galois, GraphX) have
been developed in recent years to address these needs.

Although such specialized graph data management systems have made significant advances in storing
and analyzing graph-structured data, a large fraction of the data of interest initially resides
in relational database systems (or similar structured storage systems like key-value stores, with some sort of schema); this will likely continue to be the case for a
variety of reasons including the maturity of RDBMSs, their support for transactions and SQL queries, and to some degree,
inertia.
Relational databases \cutforrevision{, as their name suggests,}
typically include many useful relationships between entities, and can contain many hidden, interesting graphs.
For example, consider the familiar DBLP dataset, where a user may want to construct a graph with the {\em authors} as the nodes. However, there are many
ways to define the {\em edges}; e.g., we may create an edge between two authors: (1) if they co-authored a paper, or (2) if they co-authored a paper \textit{recently}, or (3) if they co-authored multiple
papers together, or (4) if they co-authored a paper with very few additional authors (indicating a
true collaboration), or (5) if they attended the same conference, and so on.
Some of these graphs might be too sparse or too disconnected to yield useful insights, while others may exhibit high density or noise; however, many
of these graphs may result in different types of interesting insights. It is also often interesting to juxtapose and compare graphs constructed over
different time periods (i.e., {\em temporal graph analytics}). There are many other graphs that are possibly of
interest here, e.g., the bipartite {\em author-publication} or {\em author-conference} graphs -- identifying potentially interesting graphs
itself may be difficult for large schemas with 100s of tables.

Currently a user who wants to explore
such structures in an existing database is forced to: (a) manually formulate the right SQL queries to extract relevant data (which may not complete because of the space explosion discussed below),
(b) write scripts to convert the results into the format required by some graph database system or computation framework, (c) load the data into it,
and then (d) write and execute the graph algorithms on the loaded graphs. This is a costly,
labor-intensive, and cumbersome process, and poses a high barrier to leveraging graph analytics on these datasets. This is especially
a problem given the large numbers of entity types present in most real-world datasets and a myriad of potential graphs that could
be defined over those.

\begin{table}[t]
  \small
  \begin{center}
\begin{tabular}{ |c|c|c|c| }
\hline

\textbf{Graph} & \textbf{Representation} & \textbf{Edges} & \textbf{Extraction Time (s)} \\
\hline
\multirow{2}{5em}{DBLP\\$10M$ rows}
& Condensed & \num[group-separator={,}]{17147302} & 105.552\\
& Full Graph & \num[group-separator={,}]{86190578} & > 1200.000\\
\hline
\multirow{2}{5em}{IMDB $4.7M$ rows}
& Condensed & \num[group-separator={,}]{8437792} & 108.647\\
& Full Graph & \num[group-separator={,}]{33066098} & 687.223\\
\hline
\multirow{2}{5em}{TPCH $765K$ rows}
& Condensed & \num[group-separator={,}]{52850} & 15.52\\
& Full Graph & \num[group-separator={,}]{99990000} & > 1200.000\\
\hline
\multirow{2}{5em}{UNIV\\$32K$ rows}
& Condensed & \num[group-separator={,}]{60000} & 0.033\\
& Full Graph & \num[group-separator={,}]{3592176} & 82.042\\
\hline
\end{tabular}
\caption{Extracting graphs in \graphgen using our \textit{condensed} representation (C-DUP) vs extracting the full graph (EXP).
\graphgen enables scalable extraction and analysis on graphs that may not fit in memory.
IMDB: Co-actors graph (on a subset of data), DBLP: Co-authors graph, TPCH: Connect customers who buy the same product, UNIV: Connect students who have taken the same course (synthetic,  from \emph{http://db-book.com})}
\label{tab:benefits}
\end{center}
\vspace{-20pt}
\end{table}

\begin{figure}[t]
\begin{center}
\includegraphics[width=0.5\textwidth]{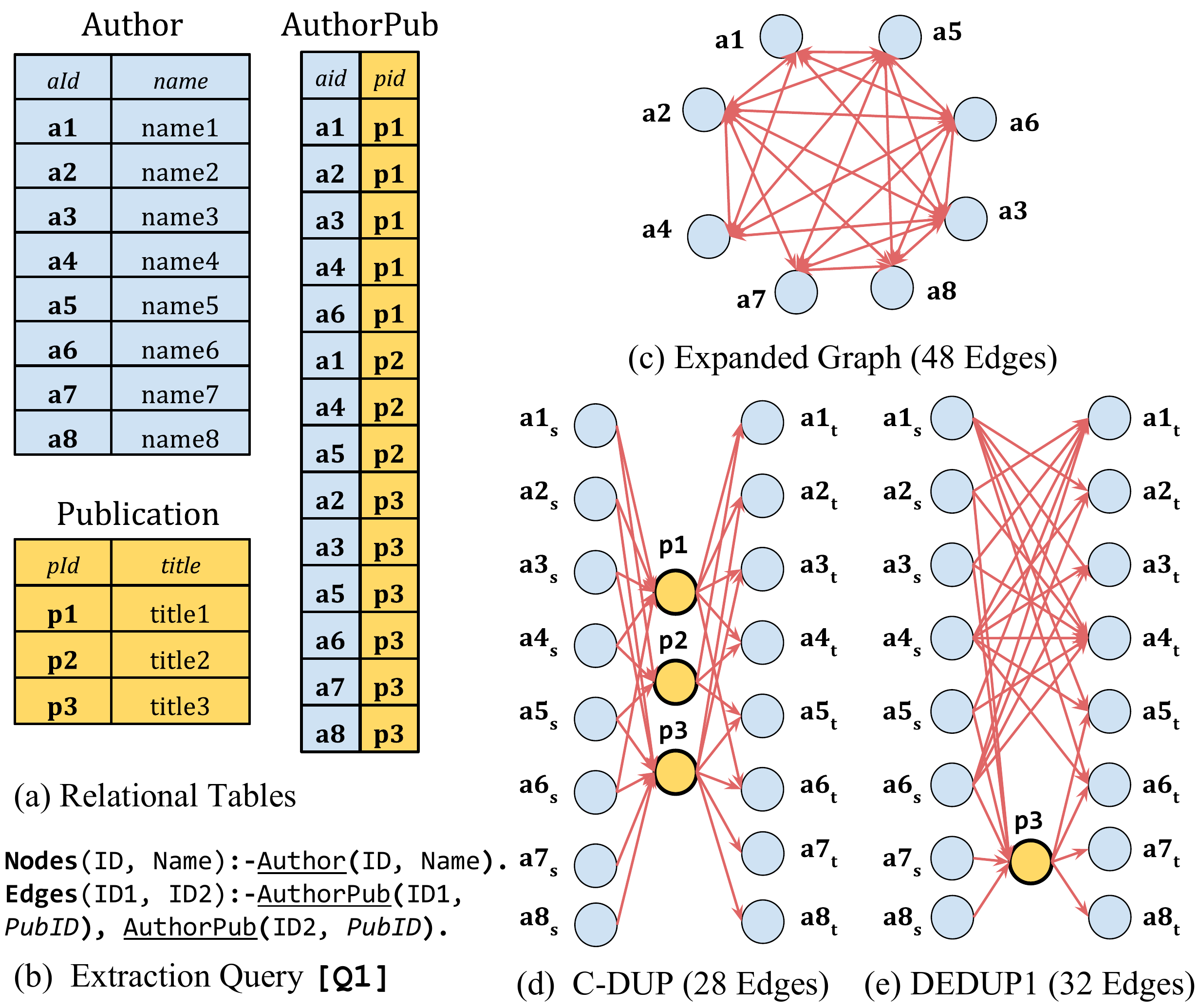}
\caption{Key concepts of \graphgen. For C-DUP and DEDUP-1, the author nodes are shown twice (with subscripts ${\_}_s$ and ${\_}_t$) to avoid clutter (by separating the in-edges and out-edges);
physically they are \textit{not} stored separately.}
\label{fig:example}
\end{center}
\end{figure}

We are  building a system, called \graphgen, with the goal to make it easy for users to extract a variety of different types of graphs from a
relational database\footnote{\small Although \graphgen currently only supports PostgreSQL, it requires only basic SQL support from the underlying storage engine, and could simply {\em scan} the tables if needed.}, and
execute graph analysis tasks or algorithms over them in memory. \graphgen supports an expressive Domain Specific Language (DSL), based on {\em
Datalog} \cite{abiteboul1995foundations}, allowing users to specify a single graph or a collection of graphs to be extracted from the
relational database (in essence, as {\em views} on the database tables). 
    \graphgen uses a translation layer to generate the appropriate SQL queries to be issued to the database, and creates an efficient in-memory representation of the graph that is handed off to the user program or analytics task. \graphgen supports a general-purpose Java
Graph API as well as the standard {\em vertex-centric API} for specifying analysis tasks like PageRank.
Figure \ref{fig:example} shows a toy DBLP-like dataset, and the query that specifies a ``co-authors'' graph to be constructed. Figure \ref{fig:example}c shows the requested co-authors graph (\graphgen naturally extracts {\em directed graphs}, and undirected graphs are represented using {\em bidirectional} edges).

\ul{The main scalability challenge in extracting graphs from relational tables is that: the graph that the user is interested in analyzing may be too
    large to extract and represent in memory, even if the underlying relational data is small}.
There is a space explosion because of the types of large-output\footnote{\small We use this term instead of ``selectivity'' terms to avoid confusion.} joins that are often needed when constructing these graphs.
Table \ref{tab:benefits} shows several examples of this phenomenon.
On the DBLP dataset restricted
to journals and conferences, there are approximately 1.6 million authors, 3 million publications, and 8.6 million author-publication relationships; the co-authors
graph on that dataset contained 86 million edges, and required more than half an hour to extract on a laptop. The {\em condensed} representation that
we advocate in this paper is much more efficient both in terms of the memory requirements and the extraction times.
The DBLP dataset is, in some sense, a best-case scenario since the average number of authors per publication is relatively small. Constructing the
{\em co-actors} graph from the IMDB dataset results in a similar space explosion.
Similarly, a graph connecting pairs of customers who bought the same item in a small TPCH dataset results in a graph much larger than the input dataset.
Even on the DBLP dataset, a graph that connects authors who have papers at the same conference contains 1.8B edges. \cutforrevision{, compared to 15M edges in the condensed representation.}

In this paper, we show how to analyze such large graphs by storing and operating upon them using a novel \textit{condensed}
representation, for extraction queries that are equivalent to \ul{unions of acyclic conjunctive queries without aggregations}. The relational model
already provides a natural such condensed representation for such queries, that we call {\em C-DUP}, obtained (essentially for free) by omitting some of the
large-output joins from the query required for graph extraction. Figure \ref{fig:example}d shows an example of C-DUP for the {\em co-authors}
graph, where we create explicit nodes for the {\em pubs}, in addition to the nodes for the {\em authors}; for two authors, $u$ and $v$, there is an edge
$u \rightarrow v$, iff there is a directed path from $u_s$ to $v_t$ in C-DUP. This representation generalizes the idea of using cliques and
bicliques for graph compression~\cite{buehrer2008scalable,karande2009speeding}; however, \ul{the key challenge for us is not generating the
representation, but rather dealing with \emph{duplicate paths} between two nodes.}

In Figure~\ref{fig:example}, we can see such a duplication for the edge $a1 \rightarrow a4$ since they are connected through both $p1$ and $p2$ . Such duplication prevents us from operating on this
condensed representation directly.
We develop a suite of different in-memory representations for this condensed graph that paired with a series of ``deduplication'' algorithms, leverage a variety of techniques for dealing with the problem of duplicate edges \cutforrevision{and ensure only \textit{a single edge} between any pair of vertices} (one of which, called DEDUP-1, is shown in Figure \ref{fig:example}e). We discuss the pros and cons of each representation and
present a formal analysis comparing their trade-offs. We also present an extensive experimental evaluation, using several real and synthetic datasets.

Key contributions of this paper include:
\begin{list}{$\bullet$}{\leftmargin 0.25in \topsep 2pt \itemsep -1pt}
\item A high-level declarative DSL based on Datalog for intuitively specifying graph extraction queries.
\item A general framework for extracting a condensed representation (with duplicates) for acyclic, aggregation-free extraction queries over arbitrary
relational schemas.
\item A suite of in-memory representations to handle the duplication in the condensed representation.
\item A systematic analysis of the benefits and trade-offs offered by extracting and operating on these representations. 
\item Novel techniques for ``deduplicating'' these condensed graphs. 
\item The first end-to-end system for enabling analytics on graphs that exist within purely relational datasets, efficiently, and without requiring complex ETL.
\end{list}

\noindent
{\bf Outline:} We begin with a brief discussion of how related work has leveraged relational databases for graph analytics in the past (Section \ref{sec:related}). We then present an overview of \graphgen, briefly describe the graph extraction DSL,
and discuss how \graphgen decides when the condensed representation should be extracted instead of the full graph (Section \ref{sec:overview}). We then
discuss the different in-memory representations (Section \ref{sec:inmemory}) and present a series of deduplication algorithms (Section \ref{sec:deduplication}). Finally, we present a comprehensive experimental evaluation (Section \ref{sec:experiments}).

\section{Related Work}
\label{sec:related}
\topic{Graph data management systems}
There has been much work on graph data management over the years, both on graph databases~\cite{angles2008survey} and graph analytics systems~\cite{dayansurvey}. Unlike graph databases, our
work targets the scenario where the data resides in an RDBMS and cannot be migrated to a graph database. The work on graph analytics systems
is largely orthogonal and complementary, as
our techniques can be used for efficient
extraction and in-memory representation in these systems, as we further discuss in Section \ref{sec:giraph}.
Several distributed graph analytics systems have adopted high-level
declarative interfaces based on Datalog~\cite{yang2016husky,seo2013socialite,gao2014glog,pregelix}. Our use of Datalog is currently restricted
to specifying which graphs to extract (in particular, we do not allow recursion). 
Combining declarative graph extraction ala \graphgen, and high-level graph analytics frameworks proposed
in that prior work, is a rich area for future work. 

Several recent works have considered how to efficiently migrate a relational database to a graph
database~\cite{relationalToGraphDBpaper,lee2015table2graph,graphbuilder}, but that work typically focuses
on creating the expanded graph most efficiently, and doesn't consider the possibility of generating a condensed representation.
Prior work on translating schemas from one data model to another has considered more complex translation problems~\cite{atzeni2008model,deutsch1999storing};
for us, the translation (through our DSL) itself is well-defined and straightforward, but the execution
of the translation query and avoiding the generation of the final result, are the main challenges.

\cutforrevision{
There also exist systems for migrating a relational database to a graph database by using the relational schema to reason about the graph
structure~\cite{relationalToGraphDBpaper}.  Table2Graph~\cite{lee2015table2graph} is built towards extracting large graphs
from relational databases using MapReduce jobs, while de-coupling the execution of the required join operations from the RDBMS.
In Table2Graph users need to provide a set of descriptive XML files that
specify the exact mappings for nodes, edges, properties and labels. Similarly,
GraphBuilder~\cite{graphbuilder} is a MapReduce-based framework for extracting
graphs from unstructured data through user-defined Java functions for node and
edge specifications. GLog~\cite{gao2014glog} is a declarative graph analysis
language based on Datalog which is evaluated into MapReduce code towards
efficient graph analytics in a distributed setting. Again, the underlying data
model they use (Relational-Graph tables) assumes that the complete graph to be
analyzed \textit{explicitly exists} as vertices and edges tables. It's
important to note that none of the mentioned works are concerned with providing
an intuitive interface or language for the mapping and extraction of hidden
graphs from the relational schema.
Users however are typically not interested
in completely migrating their data over to a graph database if they aren't
strictly dealing with graph-centric workloads.
}

\begin{figure}[t]
\begin{center}
\includegraphics[width=0.4\textwidth]{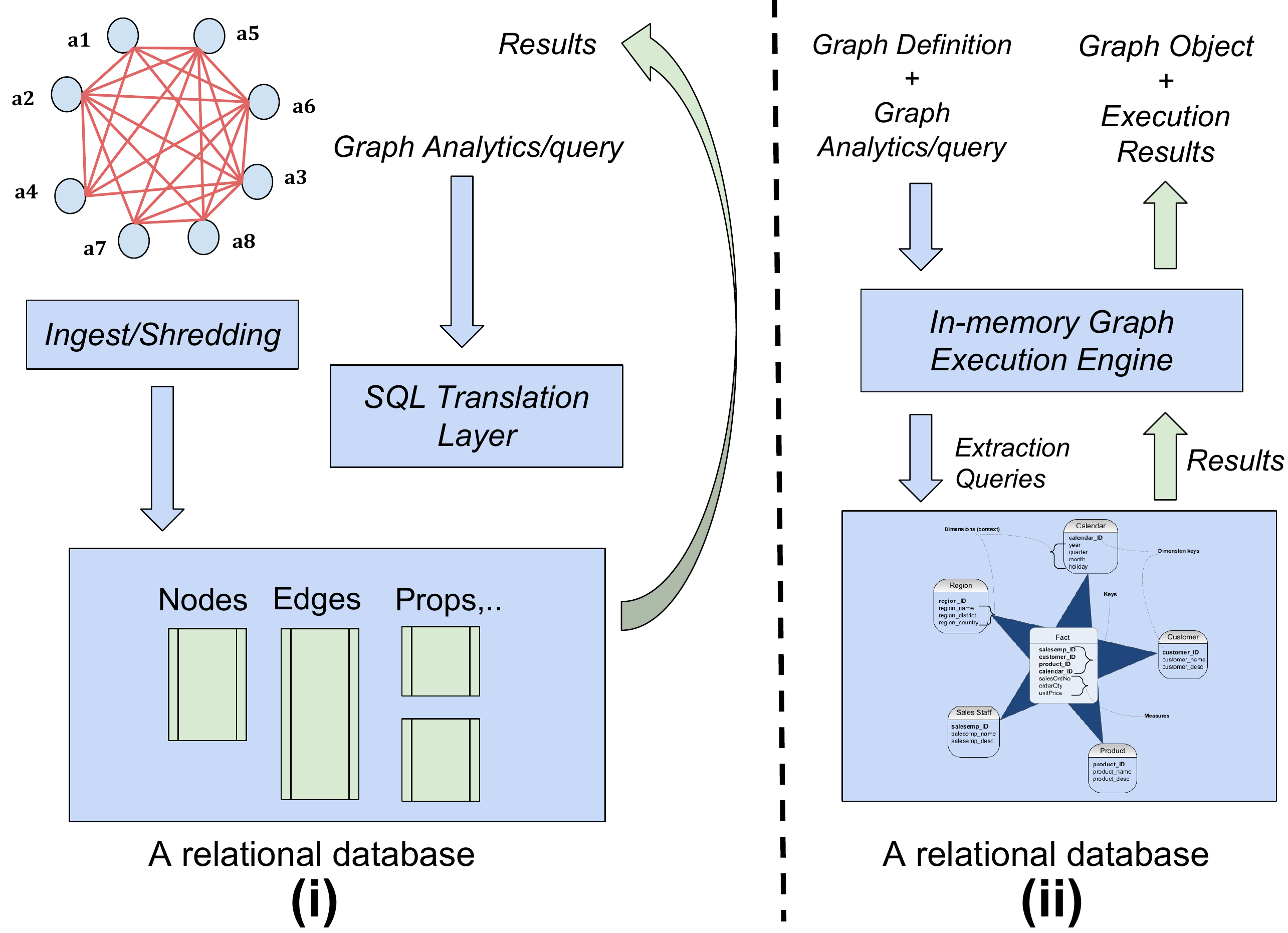}
\caption{\graphgen (right) has fundamentally different goals than recent work on using RDBMSs for graph analytics (left)}
\label{fig:comparison}
\end{center}
\end{figure}

\topic{RDBMS \& graph analytics}
Superficially the most closely related is the recent work on leveraging relational databases for graph analytics, whose aim is to show  that specialized graph
databases or analytics engines may be unnecessary. Vertexica~\cite{vertexica,jindal2014graph}, GRAIL~\cite{grail}, and SQLGraph~\cite{sun2015sqlgraph}
show how to normalize and store a graph dataset as a collection of tables in an RDBMS (i.e., how to ``shred'' graph data), and
how to map a subset of graph analysis tasks or queries to relational operations on those tables (Figure \ref{fig:comparison}).
This is similar in spirit to the earlier work on storing semi-structured or XML documents in an RDBMS~\cite{deutsch1999storing,shanmugasundaram1999relational}.
EmptyHeaded~\cite{aberger2015emptyheaded} shows how worst-case optimal join algorithms may be used for 
efficient graph querying. 
On the other hand, G-SQL~\cite{gsql} and GQ-Fast~\cite{typedgraphs} explore using graph processing engines to execute SQL queries efficiently.
However, those systems do not consider the problem of \textit{extracting} different implicit graphs from existing relational datasets, and rather choose
a relational schema for storing a given graph dataset. 
Further, those systems can typically only execute tasks (graph or XML queries) that can be mapped to SQL;
while \graphgen pushes some computation to the relational engine, most of the complex
graph algorithms are executed on a graph representation of the data in memory through a full-fledged native graph API.
This makes \graphgen suitable for complex analysis tasks like community detection, dense subgraph detection/matching,
etc., which 
require random and arbitrary access to the graph, and cannot be efficiently, if at all, executed using basic SQL.

Aster Graph Analytics~\cite{astergraph} and SAP HANA Graph Engine also
support specifying graphs within an SQL query, and applying graph algorithms on those graphs. However, the interface for specifying which graphs to extract is not very intuitive and limits the types of graphs that can be extracted. Aster only supports the vertex-centric API for writing graph algorithms. Our techniques could be
used to reduce the graph memory footprints in those systems as well.

Ringo~\cite{perez2015ringo} has somewhat similar goals to \graphgen and provides operators for converting from an in-memory relational table
representation to a graph representation. It however does not consider expensive large-output joins that are often necessary for graph extraction, or the alternate in-memory representation
optimizations that we discuss here; it instead assumes a powerful large-memory machine to deal with both issues. Ringo does include an extensive library of built-in graph algorithms in SNAP~\cite{snap}, and we do plan to support Ringo as a front-end analytics engine for \graphgen.

\topic{Factorized representation of query results} Our condensed representation is similar to the notion of ``factorized representation of query results'', where the goal is also to maintain the result of a query in a compressed form~\cite{olteanu2015size}. This prior work proposes ``schema-level factorizations'' where the decisions about how to factorize a query result are based purely on the query, the relation schemas, and the functional and multi-valued dependencies that hold on the query result. On the other hand, our approach can be seen as exploring ``data-dependent factorization''. That prior work shows that the factorization of the result of \ul{an acyclic query without projections} is linear in the size of the input database; however, for queries with projections, the storage requirements can be significantly higher. For example, for the query that generates the {\em co-authors} graph, the schema-level approach entails expanding the graph and may require quadratic storage in the worst case (cf. Appendix~\ref{app:factorized}).

Recent work on worst-case optimal joins~\cite{ngo2012worst,DBLP:conf/icdt/Veldhuizen14} shows how to avoid large intermediate results during execution of multi-way join queries; we plan to integrate those techniques into our system as we generalize our work to allow cyclic extraction queries. However, for the class of queries considered in this paper (i.e., where the edges are generated using a union of acyclic queries), those techniques do not provide any benefits over the classic Yannakakis algorithm~\cite{yannakakis}. \ul{Our challenge is that the final query result itself is too large}.

\topic{Graph Compression} There has been much work on graph, RDF, and XML compression, which can be roughly classified into~\cite{maneth2015survey}:
(a) {\em succinct representations}, where the goal is to encode the graph with as few bits as possible~\cite{boldi2004webgraph,apostolico2009graph,brisaboa2009k2,shun2015smaller};
(b) {\em structural compression}, where the graph structure is analyzed and changed to reduce the size of the graph~\cite{feder,buehrer2008scalable,asano2008efficient,buneman2003path,koch2006processing};
and (c) {\em lossy compression}, which aim to keep only sufficient information to answer specific classes of queries~\cite{fan2012query}. Our approach
is complementary to the work on succinct representations and lossy compression, and can be seen as a form of structural compression. Perhaps the most closely related work is VMiner~\cite{buehrer2008scalable}, which identifies and exploits
{\em bi-cliques} in the input graph to compress it losslessly (see Section \ref{sec:vminer} for more details).
In a recent work, Maneth and Peternek~\cite{manethcompressing} present a technique to compress a graph through detecting
repeating substructures; however, as with many of the graph or XML compression techniques, only certain types of queries can be executed
against the compressed representation.
The main difference from any of that work and our work is that: \ul{those techniques require us to first expand the graph before
compressing it}, i.e., they cannot operate on the implicit representation of the graph in the relational database; our approach aims to
avoid the expansion step itself. Our approach is better able to utilize the structure in the data, which the expansion will
remove (as our experimental results comparing to VMiner show). \ul{Further, we also support arbitrary graph operations on the compressed
representations; a necessity for a general-purpose graph engine.}

An initial prototype of the \graphgen system was recently demonstrated~\cite{demo}, where the primary focus
was on automatically proposing and extracting hidden graphs given a relational schema.
This paper provides an in-depth description of the techniques and algorithms for efficient graph extraction
as well as a comprehensive experimental evaluation of the trade-offs therein.

\vspace{5pt}
\section{System Overview}
\label{sec:overview}
We begin with a brief description of the key components of \graphgen, and how data flows through them. We then sketch our Datalog-based DSL for specifying graph
extraction jobs,
and APIs provided to the users after a graph has been loaded into memory.

\subsection{System Architecture}
\label{sec:arch}
\begin{figure}[t]
\begin{center}
\includegraphics[width=0.4\textwidth]{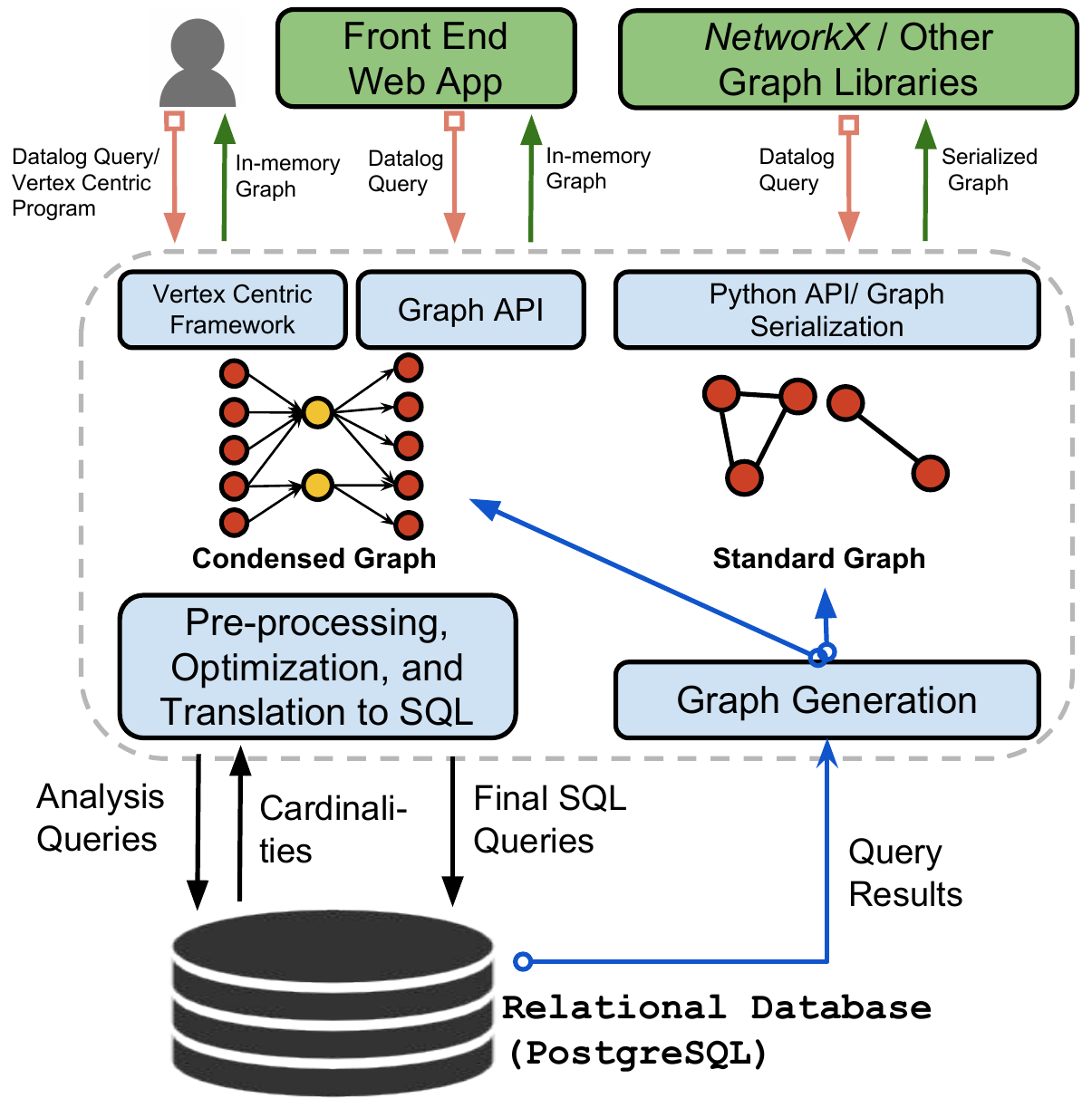}
\caption{\graphgen Overview}
\label{fig:arch}
\end{center}
\end{figure}

The inner workings of \graphgen and the components that orchestrate its functionality are demonstrated in Figure \ref{fig:arch}.
The cornerstone of the system is an abstraction layer that sits atop an RDBMS, accepts a graph extraction task, and constructs the queried graph(s) in memory, which can then be analyzed by a user program. The graph extraction task is
expressed using a Datalog-like DSL, where the user specifies how to construct the nodes and the edges of the graph (in essence, as {\em views} over the underlying tables).
This specification is parsed by a custom parser, which then analyzes the \textit{selectivities} of the joins required to construct the graph by using the statistics in the system catalog. This analysis is used to  
decide whether to hand over the partial or complete edge creation task to the database, or to skip some of the joins and load the implicit edges in memory in a condensed representation (Section \ref{sec:extraction}).

The system then builds one or more batches of SQL queries, where each batch defines one or more of the graphs that need to be extracted.
We aim to ensure that the total size of the graphs constructed in a single batch (in our memory-efficient representation) is less than the total amount of memory available, so that the graphs can be analyzed in memory. The queries are executed in sequence, and the output graph object(s) is (are) handed to the user program. A major focus of this work is to enable analysis on \textit{very large} graphs that  would typically not fit in memory.

After extraction, users can (a) operate \textit{directly} upon any portion of the graph using the Java Graph API, (b) define and run multi-threaded
\textit{vertex-centric} programs on it, (c) \textit{visually} explore the graph through our front-end web application, or (d) \textit{serialize} the
graph onto disk (in its expanded representation) in a standardized format, so that it can be further analyzed using any specialized graph processing
framework or graph library (e.g., NetworkX).

\subsection{Datalog-Based DSL}
Datalog \cutforrevision{has seen a revival in the recent years, and} has been increasingly used for expressing data analytics workflows, and especially graph analysis tasks~\cite{green2012logicblox,seo2013socialite,gao2014glog}.
The main reason for its emergence lies in its elegance for naturally expressing recursive queries, but also in its overall intuitive and simple syntax choices.
Our domain specific language (DSL) is based on a \ul{limited non-recursive subset of Datalog, augmented with looping and aggregation constructs}; in essence, our DSL allows users to
intuitively and succinctly specify \ul{nodes and edges of the \emph{target graph} as \emph{views} over the underlying database tables}. We note that our
goal is \textbf{not} to specify a graph \textit{algorithm} itself using Datalog (like Socialite~\cite{seo2013socialite}); however we do plan to explore this avenue in future work, enabling users to specify graph queries or analysis tasks using Datalog together with the graph extraction query.

Our DSL uses two special keywords:

\texttt{Nodes} and \texttt{Edges}.
A graph specification comprises of at least one \texttt{Nodes(ID, ...)} statement, followed by at least one \texttt{Edges(ID1, ID2, ...)} statement.
The first attribute of \texttt{Nodes} and first two attributes of \texttt{Edges} are assumed to denote unique \texttt{ID}s. Any additional
attributes are converted into {\em properties} (e.g., in \texttt{\textbf{[Q2]}}, the nodes of the graph will have a property {\em Name}).
Users may specify multiple \texttt{Nodes} and \texttt{Edges} statements in order to extract, say, \textit{heterogeneous} graphs with multiple
different types of vertices and edges.

Figure \ref{fig:queries} demonstrates several extraction queries. \cutforrevision{In each one of these queries, a set of  \textit{common attributes}
represents an equi-join between their respective relations.} An extraction task can contain any number of joins; e.g. \texttt{\textbf{[Q1]}} in Figure
\ref{fig:example}, only requires a single join (in this case a self-join on the \texttt{AuthorPub} table), while \texttt{\textbf{[Q2]}} as shown in
Figure~\ref{fig:query_3-4}a would require a total of 3 joins, some of which (in this case Orders(\textbf{order\_key1}, ID1) $\Join$ LineItem(\textbf{order\_key1}, part\_key), and
Orders(\textbf{order\_key2}, ID2) $\Join$ LineItem(\textbf{order\_key2}, part\_key)) will be handed off to the database since they are small-output key-foreign key joins.
The extraction query \texttt{\textbf{[Q3]}} extracts a bi-partite (heterogeneous) directed graph between instructors and students who took their courses (Figure~\ref{fig:query_3-4}b).

The \graphgen DSL naturally generates {\em directed} graphs, and undirected graphs are represented using {\em bidirectional} edges.
The typical workflow for a user when writing an extraction query would be to initially inspect the database schema, figure out which relations are relevant to the graph they are interested in exploring, and then choose which attributes in those relations would connect the defined entities in the desired way. 
We generally assume that the user is knowledgeable about the different entities and relationships existent in the database, and is able to formulate such queries. We have
built a visualization tool that allows users to discover and extract potential graphs in an interactive manner~\cite{demo}, however that is not the focus of this paper. 

\begin{figure}[t]
\scriptsize
\begin{lstlisting}[breaklines,basicstyle=\ttfamily]
[Q2] Nodes(ID, Name) :- Customer(ID, Name).
     Edges(ID1, ID2) :- Orders(order_key1, ID1),LineItem(order_key1, part_key), Orders(order_key2, ID2),LineItem(order_key2,part_key).

[Q3] Nodes(ID, Name) :- Instructor(ID, Name).
     Nodes(ID, Name) :- Student(ID, Name).
     Edges(ID1, ID2) :- TaughtCourse(ID1, courseId), TookCourse(ID2, courseId)

\end{lstlisting}
\caption{Graph Extraction Query Examples (cf. Fig.~\ref{fig:example} for \textbf{\texttt{[Q1]}})}
\vspace{-4pt}
\label{fig:queries}
\end{figure}

\subsection{Parsing and Translation}
\label{sec:translation}
\cutforrevision{
The first step towards communicating the user defined graph extraction to the system is the parsing of the Datalog query and translation into the appropriate SQL.
We have built a custom parser for the DSL described above using the ANTLR~\cite{ANTLR} parser generator. The parser is then used to create the Abstract
Syntax Tree (AST) of the query which is in turn used for translation into SQL. \graphgen evaluates programs in our DSL and translates them into SQL, line at a time. Connections between the lines of code loosely exist (e.g., code below a \texttt{For} defines a multiple ego-graph query, and translation is done accordingly), and are maintained throughout the execution of the code from one statement to the next.

The translation itself requires a full walk of the AST, during which the system gathers information about the statement, loads the appropriate
statistics for each involved relation from the database and creates a \textit{translation plan} based on the information gathered. Lastly, the
generation of the final SQL queries is actually triggered upon exiting the AST walk and is based on this translation plan.}
The first step is to parse the Datalog query, and create a set of SQL queries to execute against the database.
The specifics depend on the nature of the {\em Edges} statement(s).

\vspace{2pt}
\noindent\ul{\bf Case 1:} Each of the {\em Edges} statements corresponds to an acyclic, aggregation-free query. In that case, we may load a condensed
representation of the graph into memory (Section \ref{sec:extraction}).

\vspace{2pt}
\noindent\ul{\bf Case 2:} At least one \textit{Edges} statement violates the above condition, in which case we create a single SQL statement to construct
the edges and execute it to load the expanded graph into memory.

The rest of this paper focuses on Case 1, which corresponds to the edges being constructed using a {\em union of acyclic conjunctive queries}, and covers
many natural graph extraction tasks, including all the examples discussed so far. Even for this class of queries, extracting and operating upon the graph in
a condensed form is computationally challenging. As we discussed earlier, the recent work on factorization of query results~\cite{olteanu2015size} also has
the same goal; however, because of the projections in graph extraction queries, their techniques will expand the graph (Appendix~\ref{app:factorized}).
%
In future work, we plan to generalize our ideas to other classes of queries in Case 2, by drawing upon that prior work as well as the recent work on worst-case optimal joins; although the latter line of work enables one to avoid generating intermediate results, our focus is on avoiding generating the final result itself.

\subsection{Analyzing the Extracted Graphs}
\label{subs:analyzing}
The most efficient means to utilize \graphgen is to directly operate on the graph either using our native Java Graph API, or through a {\em vertex-centric API} that we provide. Both of these have been implemented to operate on all the in-memory (condensed or otherwise) representations that we present in Section \ref{sec:inmemory}.

\topicul{Basic Data Structure}
The basic data structure that we use for storing the graphs is a variant of traditional Compressed Sparse Row (CSR) representation~\cite{bulucc2009parallel}.
Briefly, for each node, we maintain two {\em mutable} \texttt{ArrayLists} for its in-coming and out-going edges.
We use Java \texttt{ArrayLists} instead of linked lists for space efficiency;
however, that makes {\em vertex deletions} more expensive because those require rebuilding of the entire index of vertices.
We therefore implement a lazy deletion mechanism where vertices are initially
only removed from the index, thus logically removing them from the graph, and are then physically removed from the vertices list, in \textit{batch}, at a later point in time. This way only a single re-building of the vertices index is required after a batch removal.

\topicul{Java API} All of our in-memory representations implement a simple graph API, consisting of the following 7 operations:
\begin{itemize}[noitemsep,nolistsep]
\item {\tt getVertices():} This function returns an iterator over all the vertices in the graph.
\item {\tt getNeighbors(v):} For a vertex $v$, this function returns an iterator over the neighbors of $v$, which itself supports
the standard {\tt hasNext()} and {\tt next()} functions. If a list of neighbors is desired (rather than an iterator), it can be retrieved using {\tt getNeighbors(v).toList}.
\item {\tt existsEdge(v, u):} Returns {\em true} if there is an edge between the two vertices.
\item {\tt addEdge(v, u), deleteEdge(v, u), addVer-}\\{\tt tex(v), deleteVertex(v):} These allow for manipulating the graphs by adding or removing edges or vertices.
\end{itemize}
The {\tt Vertex} class also supports setting or retrieving properties associated with a vertex.

\topicul{Vertex-centric API} The vertex-centric conceptual model has been extensively used in the past to express complex graph algorithms by following the  ``think-like-a-vertex'' methodology in designing these algorithms. We have implemented a simple, multi-threaded variant of the \textit{vertex-centric framework} in \graphgen that allows users to implement a \textsc{compute} function and then execute that against the extracted graph regardless of its in-memory representation. The framework is based on a \texttt{VertexCentric} object which coordinates the multi-threaded execution of the \texttt{compute()} function for each job. The coordinator object splits the graph's nodes into chunks depending on the number of cores in the machine, and distributes the load evenly across all cores. It also keeps track of the current superstep, monitors the execution and triggers a termination event when all vertices have voted to a halt. Users simply need to implement the \texttt{Executor} interface which contains a single method definition for \texttt{compute()}, instantiate their executor and call the \texttt{run()} method of the VertexCentric coordinator object with the Executor object as input. The implementation of message passing we've adopted is similar to the \emph{gather-apply-scatter (GAS)} model used in GraphLab~\cite{graphLab} in which nodes communicate by directly accessing their neighbors' data, thus avoiding the overhead of explicitly storing messages in some intermediary data structure.

\begin{figure}[t]
  \begin{center}
  \includegraphics[width=0.50\textwidth]{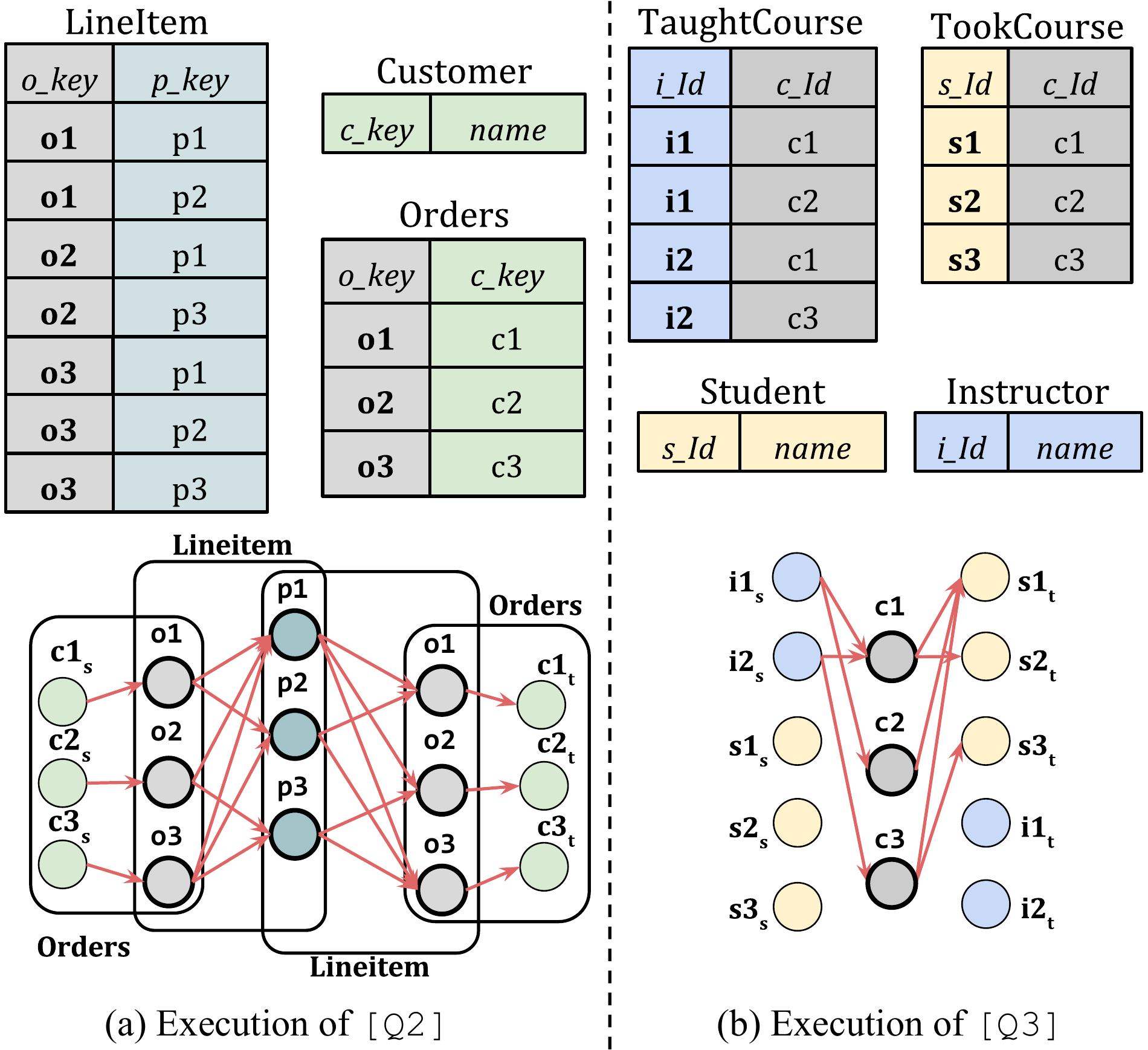}
  \vspace{-5pt}
  \caption{Extraction examples: (a) Multi-layered condensed representation, (b) extracting a heterogeneous bipartite graph (we only list the
schemas for some of the tables, and omit tuples for clarity)}
  \vspace{-5pt}
  \label{fig:query_3-4}
  \end{center}
\end{figure}

\topicul{External Libraries} \graphgen can also be used through a  library called \textit{graphgenpy}\footnote{\small \url{http://konstantinosx.github.io/graphgen-project/}},
a Python wrapper over \graphgen allowing users to run queries in our DSL through simple Python
scripts and serialize the resulting graphs in a standard graph format, thus opening up analysis to any graph computation framework or library (Section~\ref{sec:giraph}). A similar workflow was used in the implementation of our front-end web application~\cite{demo} through which users can visually explore the graphs that
exist within their relational schema. 

\section{In-Memory Representation and Task Execution}
\label{sec:inmemory}

The key efficiency challenge with extracting graphs from relational databases is that: in most cases, \cutforrevision{because of the normalized nature
of relational schemas,} queries for extracting explicit relationships (i.e., {\em edges}) between entities from relational datasets
(i.e., {\em nodes}) requires expensive non-key (large-output) joins. Because of this, the extracted graph may be much
larger than the input size itself (cf. Table~\ref{tab:benefits}). Instead, we propose maintaining and operating upon the extracted graph
in a {\em condensed} fashion. We begin with describing a novel condensed representation that we use and discuss why it is ideally suited for
this purpose; we also discuss the {\em duplication} issue with this representation. We then present a general algorithm for constructing
such a condensed representation for a graph extraction query over an arbitrary schema, that guarantees that the condensed representation requires
at most as much memory as loading all the underlying tables in the worst case. The condensed representation and the algorithm
can both handle any non-recursive, acyclic graph extraction query over an arbitrary schema.

We then propose a series of in-memory variations of the basic condensed representation that handle the duplication issue through uniquely
characterized approaches. 
These representations are products of a single-run preprocessing phase on top of the condensed representation using an array of algorithms described in Section \ref{sec:deduplication}. We also expand on the natural trade-offs associated with storing and operating on each in-memory representation.

\subsection{Condensed Representation \& Duplication}
\label{subs:condensed_representation}
The idea of compressing graphs through identifying specific types of structures has been around for a long time~\cite{feder,buehrer2008scalable}.
As discussed in Section \ref{sec:related}, those prior techniques are not directly applicable here since they require the input graph to be
in an expanded form before compressing.
Instead, we propose a novel condensed representation, called C-DUP, that is effectively free to construct from the database and
requires less memory to load. 
Given a graph extraction query, let $G(V, E)$ denote the output {\bf expanded} graph; for clarity of exposition, we assume that $G$ is a directed
graph. Since in the C-DUP representation there are only edges between \textit{real} nodes and \textit{virtual} nodes, we denote vertices with subscript $\__s$(source) to describe the vertices that have out-edges to virtual nodes, and vertices with subscript $\__t$(target) for vertices that have in-edges from virtual nodes. We say $G_C(V', E')$ is an \ul{\emph{equivalent C-DUP representation}} if and only if:\\[2pt]
(1) for every node $u \in V$, there are two nodes $u_s, u_t \in V'$ -- the remaining nodes in $V'$ are called \ul{\em virtual nodes};\\[2pt]
(2) $G_C$ is a directed {\em acylic} graph, i.e., it has no directed cycle; \\[2pt]
(3) in $G_C$, there are no incoming edges to $u_s \forall u \in V$ and no outgoing edges from $u_t \forall u \in V$; \\[2pt]
(4) for every edge $\langle u \rightarrow v\rangle  \in E$, there is at least one directed path from $u_s$ to $v_t$ in $G_C$.

Figure \ref{fig:query_3-4} shows two examples of such condensed graphs. In the second case, where a heterogeneous bipartite graph is being extracted, there
are no outgoing edges from $s1_s, s2_s, s3_s$ or incoming edges to $i1_t, i2_t$, since the output graph itself only has edges from $i\_$ nodes to $s\_$
nodes. Although we assume there are two copies of each real node in $G_C$ here, the physical representation of $G_C$ only requires one copy
(with special-case code to handle incoming and outgoing edges). There may be self-edges in the extracted graph (e.g., $c1_s \rightarrow c1_t$ in Figure \ref{fig:query_3-4}a); however, since our extraction queries are acyclic, $G_C$ itself is always a directed acyclic graph.

\topicul{Duplication Problem}
Although the C-DUP representation is easy to construct, it allows for multiple paths between $u_s$ and $v_t$, since
that's the natural output of the extraction process below.
\textit{Any} graph
algorithm whose correctness depends solely on the connectivity structure of the graph (i.e., ``duplicate-insensitive'' algorithms), can be
executed directly on top of this representation, with a potential for \textit{speedup} (e.g., connected components or breadth-first search);
the notion of {\em representation-independent graph analytics} from recent work could be used to further
increase the applicability of the C-DUP representation~\cite{hogan2014everything,chodpathumwan2015universal}.
However, this \textit{duplication} causes correctness issues on all non duplicate-insensitive graph algorithms. The duplication problem entails that programmatically, when each real node tries to iterate over its neighbors, passing through its obligatory virtual neighbors, it may encounter the same neighbor \textit{more than once}; this indicates a duplicate edge. The set of algorithms we propose in Section~\ref{sec:deduplication} are geared towards dealing with this duplication problem.

\topicul{Single-layer vs Multi-layer Condensed Graphs}
A condensed \\ graph may have one or more layers of virtual nodes (formally, a condensed graph is called {\em multi-layer} if it contains a directed
path of length > 2).  In the majority of cases, most of the joins involved in extracting these graphs will be simple key-foreign key joins, and
large-output joins (which require use of virtual nodes) occur relatively rarely. Although our system can handle arbitrary multi-layer graphs, we
also develop special algorithms for the common case of single-layer condensed graphs.

\subsection{Extracting a Condensed Graph}
\label{sec:extraction}
The key idea behind constructing a condensed graph is to postpone certain joins. Here we briefly sketch our algorithm for making those decisions,
extracting the graph, and postprocessing it to reduce its size.

\topic{Step 1} First, we translate the {\em Nodes} statements into SQL queries, and execute those against the database to load the nodes in memory.
In the following discussion, we assume that for every node $u$, we have two copies $u_s$ ({\em source}) and $u_t$ ({\em target}); physically we only store one
copy.

\topic{Step 2} We consider each {\em Edges} statement in turn. Recall that the output of each such statement is a set of 2-tuples (corresponding to a set of
edges between real nodes), and further that we assume the statement is acyclic and aggregation-free (cf. Section \ref{sec:translation}, Case 1). Without
loss of generality, we can represent the statement as:\\[2pt]
$Edges(ID1, ID2) :- R_1(ID1, a_1), R_2(a_1, a_2), ..., R_n(a_{n-1}, ID2)$\\[2pt] (two different
relations, $R_i$ and $R_j$, may correspond to the same database table). Generalizations to allow multi-attribute joins and selection predicates are
straightforward.

For each join $R_i(a_{i-1}, a_i) \Join_{a_i} R_{i+1}(a_i, a_{i+1})$, we retrieve the number of distinct values, $d$, for $a_i$ (the join attribute) from the
system catalog (e.g., \texttt{n\_distinct} attribute in the
\texttt{pg\_stats} table in PostgreSQL). If $|R_i||R_{i+1}| / d > 2 (|R_i| + |R_{i+1}|)$, then we consider this a {\em large-output} join and mark it so
(this formula assumes that the join attribute is uniformly distributed and may miss a large-output join and could be easily substituted with
a more sophisticated selectivity estimator).

\topic{Step 3}
We then consider each subsequence of the relations without a large-output join, construct an SQL query corresponding to it, and execute it against
the database. Let $a_l, a_m, ..., a_u$ denote the join attributes which are marked as {\em large-output}. Then, the queries we execute correspond to:\\
$res_1(ID1, a_l) :- R_1(ID1, a_1), ..., R_{l}(a_{l-1}, a_l)$, \\[1pt] $res_2(a_l, a_m) :- R_{l+1}(a_l, a_{l+1}), ..., R_{m}(a_{m-1}, a_m)$, ..., and
\\[1pt]
$res_k(a_u, ID2) :- R_{u+1}(a_u, a_{u+1}), ..., R_n(a_{n-1}, ID2)$.

\topic{Step 4} For each join attribute $attr \in \{a_l, a_m, ..., a_u\}$, we create a set of virtual nodes corresponding to all possible values
$attr$ takes.

\topic{Step 5} For $(x, y) \in res_1$, we add a directed edge from a real node to a virtual node: $x_s \rightarrow y$. For $(x, y) \in res_k$, we add a directed edge $x \rightarrow
y_t$. For all other $res_i$, for $(x, y) \in res_i$, we add an edge between two virtual nodes: $x \rightarrow y$.

\topic{Step 6 (Preprocessing)} For a virtual node, let $in$ and $out$ denote the number of incoming and outgoing edges respectively; if $in
\times out \le (in + out + 1)$, we ``expand'' this node, i.e., we remove it and add directed edges from its in-neighbors to its out-neighbors. This preprocessing
step can have a significant impact on memory consumption. We have implemented a multi-threaded version of this to exploit multi-core machines,
which resulted in several non-trivial concurrency issues. We omit a detailed discussion for lack of space. Finally, the system also computes
the number of edges in the expanded graph (this can be computed for free as a side-effect of all of our deduplication algorithms), and expands
the graph if the increase in size is small.

If the query contains multiple {\em Edges} statements, the final constructed graph would be the union of the graphs constructed for each of them.
It is easy to show that the constructed graph satisfies all the required properties listed above, that it is equivalent to the output graph, and
it occupies no more memory than loading all the input tables into memory.

In the example shown in Figure~\ref{fig:query_3-4}a, the graph specified in query \texttt{\textbf{[Q2]}} that is extracted assumes that all three of the joins involved portray low selectivity, and so we choose not to hand any of them to the database, but extract the condensed representation by instead projecting the tables in memory and creating intermediate virtual nodes for each unique value of each join condition.

\subsection{In-Memory Representations}
\label{subs:representations}
Next, we propose a series of in-memory graph representations that can be utilized to store the condensed representation mentioned above, in its \textit{deduplicated} state.
Here we discuss the representation formats and their key properties, and with a specific focus on the implementation of
the \textit{getNeighbors()} iterator, which underlies most graph algorithms. We note that, in some cases, we use the same term to both denote an in-memory representation,
as well as the algorithm for constructing that representation; e.g., we discuss the DEDUP-1 representation below and outline its key properties (e.g., it does not suffer
from duplication), and we discuss several algorithms for constructing the DEDUP-1 representation in the next section.
We note that our representations primarily explore different ways to do structural compression, and could be combined with other graph compression approaches~\cite{maneth2015survey} to further reduce the memory footprint.

\topicul{C-DUP: Condensed Duplicated Representation}
This is the representation that we initially extract from the relational database, which suffers from the edge duplication problem. We can utilize this representation as-is by employing a naive solution to deduplication, i.e., by doing deduplication \textit{on the fly} as algorithms are being executed.
Specifically, when we call \textit{getNeighbors(u)}, it starts a \ul{depth-first traversal} from $u_s$ and returns all the real nodes ($\__t$ nodes)
reachable from $u_s$; it also keeps track of which neighbors have already been seen (in a {\em hashset}) and skips over them if the neighbor is seen
again.

This is typically the most memory-efficient representation, does not require any preprocessing overhead, and is a good option for graph algorithms
that access a small fraction of the graph (e.g., if we were looking for information about a small number of specific nodes). On the other hand, due
to the required hash computations at every call, the \ul{execution penalty}  for this representation is high, especially for multi-layer graphs; it
also suffers from \ul{memory and garbage collection bottlenecks} for algorithms that require processing all the nodes in the graph. Operations like \texttt{deleteEdge()} are also quite
involved in this representation, as deletion of a logical edge may require non-trivial modifications to the virtual nodes.

\topicul{EXP: Fully Expanded Graph}
On the other end of the spectrum, we can choose to expand the graph in memory, i.e., create all direct edges between all the real nodes in the graph and remove the virtual nodes. The expanded graph typically has a much larger memory footprint than the other representations due to the large number of edges.
It is nevertheless, naturally, the most efficient representation for operating on, since iteration only requires a sequential scan over one's direct neighbors. The expanded graph is the baseline that we use to compare the performance of all other representations in terms of trading off memory with
operational complexity.

\topicul{DEDUP-1: Condensed Deduplicated Representation}
This representation format is identical to C-DUP in its use of virtual nodes, with the major difference being that it does not suffer from duplicate paths,
and thus does not require the on-the-fly deduplication used in C-DUP (i.e., {\em getNeighbors()} does not need to use the {\em hashset}).
This representation typically sits in the middle of the spectrum between EXP and C-DUP in terms of both memory efficiency and iteration performance; it usually results in a larger number of edges than C-DUP, but has reduced overhead of
neighbor iteration. The trade-offs here also include the one-time cost of removing duplication; \ul{deduplicating a graph while minimizing the number of
edges added can be shown to be NP-Hard}. Unlike the other representations discussed below, this representation maintains the simplicity of C-DUP and
can easily be serialized and used by other systems which need to simply implement a proper iterator.

\begin{figure}[t]
  \includegraphics[width=0.50\textwidth]{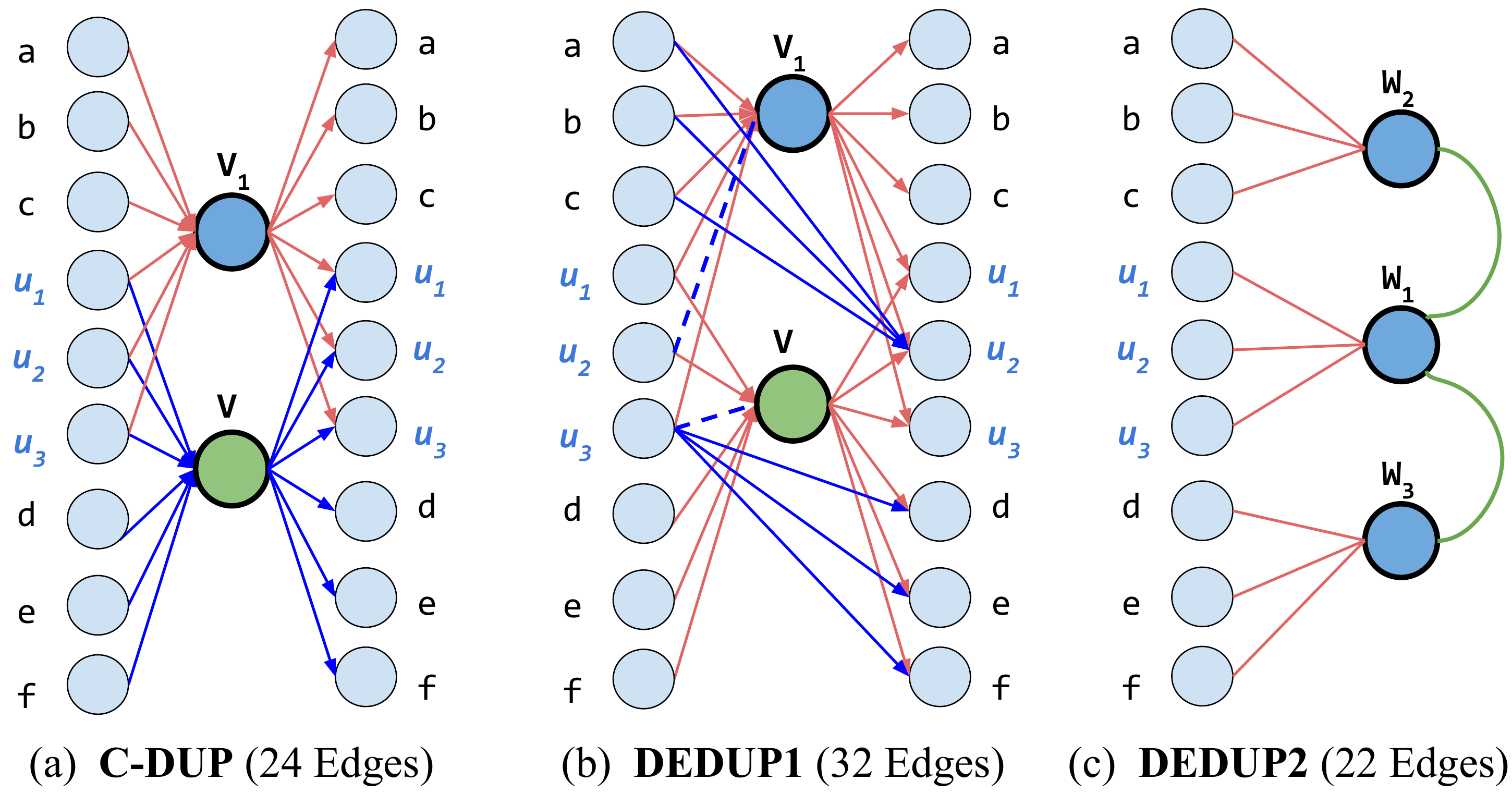}
  \vspace{-6pt}
         \caption{The resulting graph after the addition of virtual node $V$. (c) shows the resulting graph for if we added edges \textit{between} virtual nodes (we omit $\__s$ and $\__t$ subscripts since they are clear from the context).}
         \label{fig:dedup2_examples}
         \vspace{-4pt}
\end{figure}

\topicul{BITMAP: Deduplication using Bitmaps}
This representation results from applying a different kind of preprocessing based on maintaining \textit{bitmaps}, for
\textit{filtering} out duplicate paths between nodes. Specifically, a virtual node $V$ may be associated with a set of bitmaps, indexed by the IDs of
the real nodes; the size of each bitmap is equal to the number of outgoing edges from $V$. Consider a depth-first traversal starting at $u_s$ that
reaches $V$. We check to see if there is a bitmap corresponding to $u_s$; if not, we traverse each of the outgoing edges in turn. However, if there
is indeed a bitmap corresponding to $u_s$, then we consult the bitmap to decide which of the outgoing edges to skip (i.e., if the corresponding bit
is set to 1, we traverse the edge). In other words, the bitmaps are used to eliminate the possibility of reaching the same neighbor twice.

The main drawback of this representation is the memory overhead and complexity of storing these bitmaps, which also makes this representation
less \textit{portable} to systems outside \graphgen. The preprocessing required to set these bitmaps can also be quite involved as we discuss in the
next section.

\topicul{DEDUP-2: Optimization for Single-layer Symmetric Graphs} \\
This optimized representation can significantly reduce the memory requirements for dense graphs, for the special case of a single-layer, {\em symmetric} condensed graph (i.e., $\langle u_s \rightarrow v_t \rangle \implies \langle v_s \rightarrow u_t \rangle$); many graphs satisfy these conditions. In such a case,
for a virtual node $V$, if $u_s \rightarrow V$, then $V \rightarrow u_t$, and we can omit the $\__t$ nodes and associated edges.
Figure \ref{fig:dedup2_examples} illustrates an example of the same graph if we were to use all three deduplication representations. In C-DUP, we have two virtual nodes $V_1$ and $V_2$, that are both connected
to a large number of real nodes. The optimal DEDUP-1 representation (Figure \ref{fig:dedup2_examples}b) results in a substantial increase in the
number of edges, because of the large number of duplicate paths. The DEDUP-2 representation (Figure \ref{fig:dedup2_examples}c) uses {\em special}
undirected edges between virtual nodes to handle such a scenario. A real node $u$ is considered to be connected to all real nodes that it can reach through each of its direct neighboring virtual nodes $v$, \textit{as well as} the virtual nodes directly connected to $v$ (i.e. 1 hop away); e.g., node $a$ is connected to $b$ and $c$ through $W_2$, and to $u_1, u_2, u_3$ through $W_1$ (which is connected
        to $W_2$), but not to $d, e, f$ (since $W_3$ is not connected to $W_2$). This representation is required to be duplicate-free, i.e., there can be at most one such path between a pair of nodes. The DEDUP-2 representation here requires 11 undirected edges, which is just below the space requirements for C-DUP. However, for dense graphs, the benefits can be substantial (Section \ref{sec:experiments}).

Generating a good DEDUP-2 representation for a given C-DUP graph is much more intricate than generating a DEDUP-1 representation. Due to space constraints, we omit the algorithm from the main body of the paper, and present a sketch in Appendix \ref{app:dedup2}.

\begin{figure}[t]
\begin{center}
\includegraphics[width=0.48\textwidth]{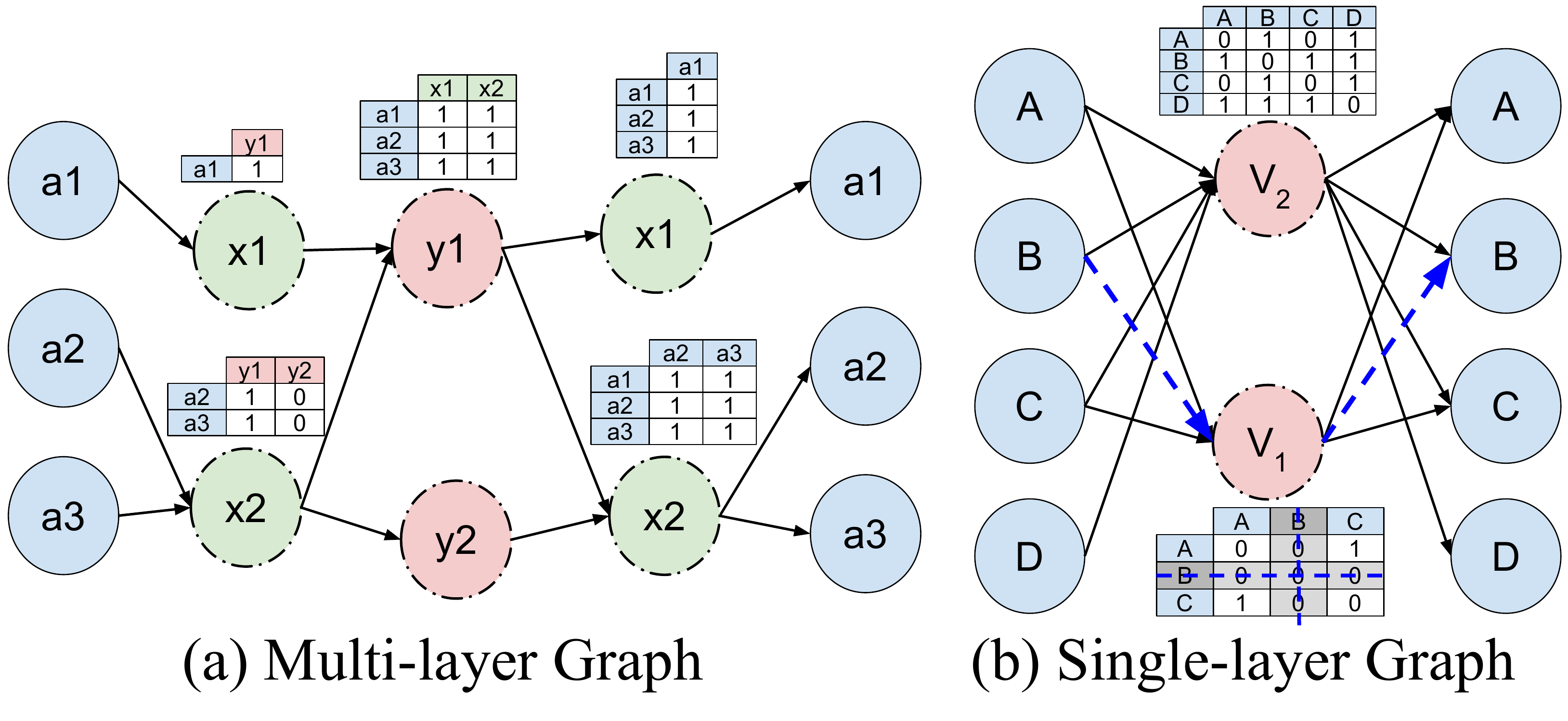}
\caption{Using \textit{BITMAP}s to handle duplication; the dotted edges (corresponding to columns or edges
with all 0s) are removed.}
\vspace{-20pt}
\label{fig:bitmap_example}
\end{center}
\end{figure}

\section{Preprocessing \& Deduplication}
\label{sec:deduplication}

In this section, we discuss a series of preprocessing and deduplication algorithms we have developed for constructing the different in-memory
representations for a given query. The input to all of these algorithms is the C-DUP representation, that has been extracted and instantiated in
memory. We first present a general preprocessing algorithm for the BITMAP representation for multi-layer condensed graphs. We then discuss a series
of optimizations for single-layer condensed graphs, including deduplication algorithms that attempt to eliminate duplication (i.e., achieve DEDUP-1
representation). In contrast with Section~\ref{subs:representations}, here we focus on describing algorithms for \textit{how} to generate the representations that we described there.   
We also describe the runtime complexity for each algorithm in which we refer to $n_r$ as the number of real nodes, $n_v$ as the number of virtual nodes, $k$ as the number of layers of virtual nodes, and $d$ as the maximum degree of any node (i.e., the maximum number of outgoing edges).

\subsection{Preprocessing for BITMAP}
\label{subs:bmp_proc}
Recall that the goal of the preprocessing phase here is to associate and initialize bitmaps with the virtual nodes to avoid visiting the same real node
twice when iterating over the out-neighbors of a given real node.
We begin with presenting a simple, naive algorithm for setting the bitmaps; we then analyze the complexity of doing so optimally and present a set cover-based greedy
algorithm.

\subsubsection{BITMAP-1 Algorithm}
\label{subs:BITMAP1}
This algorithm only associates bitmaps with the virtual nodes in the penultimate layer, i.e., with the virtual nodes that have outgoing edges to $\__t$ nodes.
We iterate over all the real nodes in turn. For each such node $u$, we initiate a depth-first traversal from $u_s$, keeping track of all the real nodes visited
during the process using a hashset, $H_u$. For each virtual node $V$ visited, we check if it is in the penultimate layer; if yes, we add a bitmap of size equal to the number of outgoing
edges from $V$. Then, for each outgoing edge $V \rightarrow v_t$, we check if $v_t \in H_u$. If so, we set the corresponding bit to 0; else, we set it to 1 and add $v_t$ to $H_u$.

This is the least computationally complex of all the algorithms, and in practice the fastest algorithm. It maintains the same number of edges as C-DUP, while adding the overhead of the bitmaps and the appropriate indexes associated with them for each virtual node.
The traversal order in which we process each real node does not matter here since the end result will always have the same number of edges as C-DUP. Changing the processing order only changes the way the {\bf set} bits are distributed among the bitmaps.

\topic{Complexity} The worst-case runtime complexity of this algorithm is $O(n_r*d^{k+1})$. Although this might seem high, we note that this is
always lower than the cost of expanding the graph.

\subsubsection{Formal Analysis}
The above algorithm, while simple, tends to initialize and maintain a large number of bitmaps. This leads us to ask the question: \ul{how can we
achieve the required deduplication while using the minimum number of bitmaps (or minimum total number of bits)?} This seemingly simple problem
unfortunately turns out to be NP-Hard, even for single-layer graphs.
In a single-layer condensed graph, let $u$ denote a real node, with edges to virtual nodes $V_1, ..., V_n$, and let $O(V_1)$ denote the set of real
nodes to which $V_1$ has outgoing edges. Then, the problem of identifying a minimum set of bitmaps to maintain is equivalent to finding a {\em set
cover} where our goal is to find a subset of $O(V_1), ..., O(V_n)$ that covers their union. Unfortunately, the set cover problem is not only NP-Hard,
      but is also known to be hard to approximate.

\subsubsection{BITMAP-2 Algorithm}
This algorithm is based on the standard {\em greedy algorithm} for set cover, which is known to achieve the best approximation ratio ($O(\log n)$)
for the problem. We describe it using the terminology above for single-layer condensed graphs. The algorithm starts by picking the virtual node $V_i$ with
the largest $|O(V_i)|$. It adds a bitmap for $u$ to $V_i$, and sets it to all 1s; all nodes in $O(V_i)$ are now considered to be {\em covered}. It then identifies the virtual node $V_j$ with the largest $|O(V_j) -
O(V_i)|$, i.e., the virtual node that connects to largest number of nodes that remain to be covered. It adds a bitmap for $u_s$ to $V_j$ and sets it
appropriately. It repeats the process until all the nodes that are reachable from $u_s$ have been covered. For the remaining virtual nodes (if any),
the edges from $u_s$ to those nodes are simply deleted since there is no reason to traverse those.

We generalize this basic algorithm to multi-layer condensed graphs by applying the same principle at each layer. Let $V^1_1, ..., V^1_n$ denote the set of
virtual nodes pointed to by $u_s$. Let $N(u_s)$ denote all the real $\__t$ nodes reachable from $u_s$. For each $V^1_i$, we count how many of the
nodes in $N(u_s)$ are reachable from $V^1_i$, and explore the virtual node with the highest such count first. At the penultimate layer, the algorithm
reduces to the single-layer algorithm described above and appropriately sets the bitmaps. We consistently keep track of how many of the nodes in
$N(u_s)$ have been covered so far, and use that for making the decisions about which bits to set. So after bitmaps have been set for all virtual
nodes reachable from $V^1_1$, if there are still nodes in $N(u_s)$ that need to be covered, we pick the virtual node $V^1_i$ that reaches the largest
number of uncovered nodes, and so on.

It's important to note that here we never delete an outgoing edge from a virtual node, since it may be needed for another real node. Instead, we use
bitmaps to stop traversing down those paths (e.g., edge $x_2 \rightarrow y_2$ in Figure \ref{fig:bitmap_example}).

Our implementation exploits multi-core parallelism, by creating equal-sized chunks of the set of real
nodes, and processing the nodes in each chunk in parallel.

\topic{Complexity} The runtime complexity of this algorithm is significantly higher than BITMAP-1 because of the need to re-compute the number of
reachable nodes after each choice, and the worst-case complexity could be as high as: $O(n_r*d^{2^k})$. In practice, $k$ is usually 1 or 2,
and the algorithm finishes reasonably quickly, especially given our parallel implementation. 

\begin{figure}[t]
\begin{center}
\vspace{-8pt}
\includegraphics[width=0.5\textwidth]{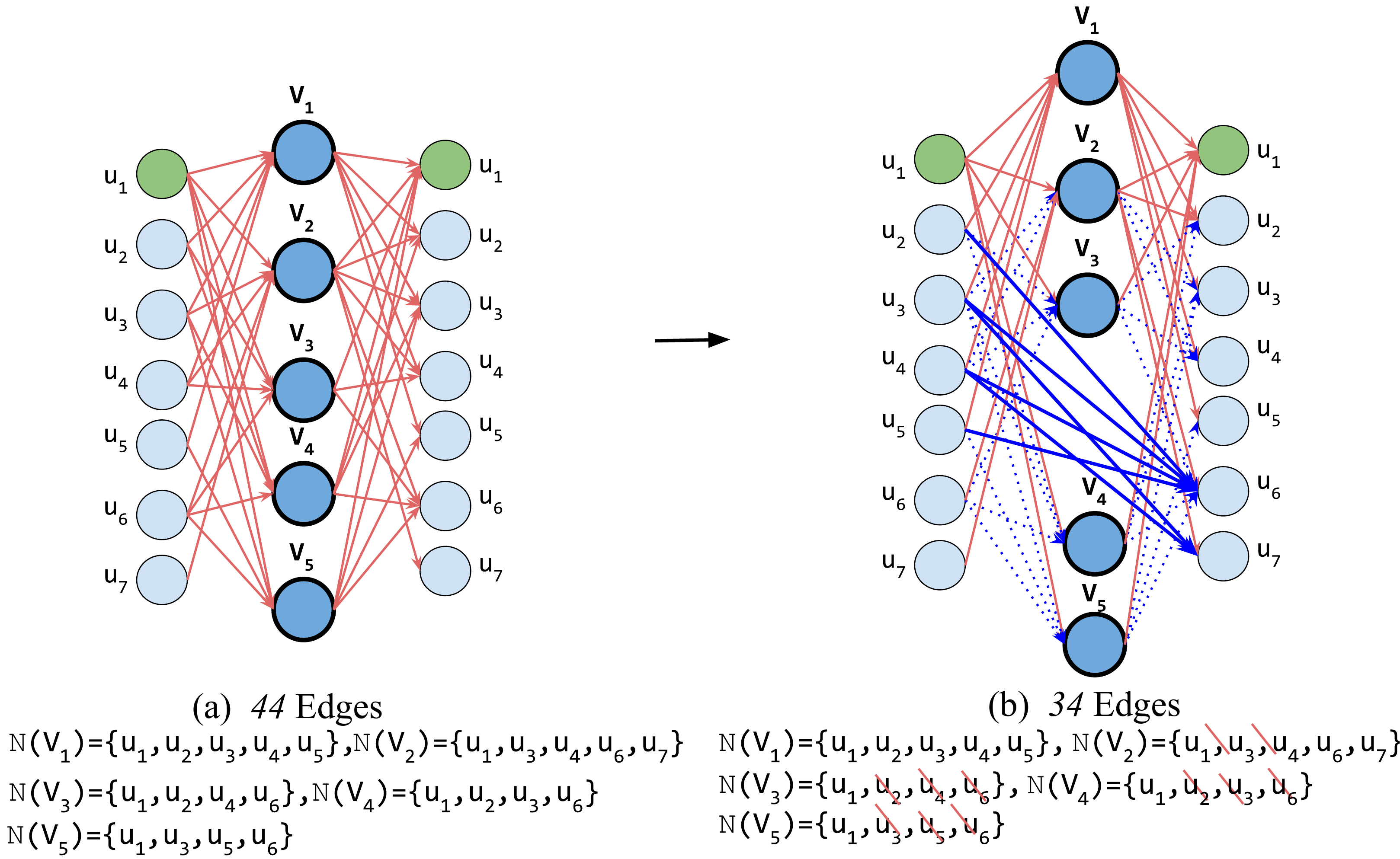}
\vspace{-5pt}
\caption{Deduplicating $u_1$ using the ``real-nodes first'' algorithm, resulting to an equivalent graph with a \textit{smaller} number of edges}
\vspace{-5pt}
\label{fig:real_nodes_first}
\end{center}
\end{figure}

\subsection{Deduplication for DEDUP-1}
    The goal with deduplication is to
modify the initial C-DUP graph to reach a state where there is at most one unique path
between any two real nodes in the graph.
We describe a series of novel algorithms for achieving this for single-layer condensed graphs, and discuss the pros and cons of using each one as well as
their effectiveness in terms of the size of the resulting graph. We briefly sketch how these algorithms can be extended to multi-layer condensed graphs; however, we leave a
detailed study of deduplication for multi-layer graphs to future work.

\subsubsection{Single-layer Condensed Graphs}
The theoretical complexity of this problem for single-layer condensed graphs is the same as the original problem considered by Feder and Motwani~\cite{feder}, which focuses on the reverse problem of \textit{compressing cliques} that exist in the expanded graph, by finding cliques and connecting all vertices in the same clique to a virtual node and removing the edges among them. Although the expanded graph is usually very large, it is still only $O(n^2)$, so
    the NP-Hardness of the deduplication problem is the same. However, those algorithms presented in \cite{feder} are not applicable here because the input representation is different, and expansion is not an option. We present four algorithms for this problem.

In the description below, for a virtual node $V$, we use $I(V)$ to denote the set of real nodes that point to $V$, and $O(V)$ to denote the real nodes that $V$ points to.

\topicul{Naive Virtual Nodes First}
\label{subs:naive_virt}
This algorithm deduplicates the graph one virtual node at a time. We start with a graph containing only the real nodes and no virtual nodes, which is trivially duplication-free. We then add the virtual nodes one at a time, always ensuring that the partial graph remains free of any duplication.

When adding a virtual node $V$: we first collect all of the virtual nodes $R_i$ such that $I(V) \cap I(R_i) \ne \phi$; these are the virtual nodes that the real nodes in $I(V)$ point to. Let this set be $R$. A $processed$ set is
also maintained which keeps track of the virtual nodes that have been added to the current partial graph.
For every virtual node $R_i \in R \cap processed$, if $|O(V) \cap O(R_i)| > 1$, we modify the virtual nodes to handle the duplication before adding $V$ to the partial graph (If there is no such $R_i$, we are done). We select a real node $r \in O(V)\cap O(R_i)$ at \textit{random}, and choose to either remove the edge
$(V \rightarrow r)$ or $(R_i \rightarrow r)$, depending on the in-degrees of the two virtual nodes. The intuition here is that, by removing the edge from the lower-degree virtual node, we have to add fewer direct edges to compensate for removal of the edge.
Suppose we remove the former ($V \rightarrow r$) edge.
We then add direct edges to $r$ from all the real nodes in $I(V)$, while checking to make sure that $r$ is not already connected to those nodes through other virtual nodes. Virtual node $V$ is then added to a $processed$ set and we consider the next virtual node.

\topic{Complexity} The runtime complexity is $O(n_v*d^4)$.

\topicul{Naive Real Nodes First}
In this approach, we consider each \textit{real node} in the graph at a time, and handle duplication between the virtual nodes it is connected to, in the order in which they appear in its neighborhood. This algorithm handles deduplication between two virtual nodes that overlap in exactly the same way as the one described above. It differs however in that it entirely handles \textit{all duplication} between a single real node's virtual neighbors before moving on to processing the next real node. As each real node is handled, its virtual nodes are added to a $processed$ set, and every new virtual node that comes in is checked for duplication against the rest of the virtual nodes in this $processed$ set. This $processed$ set is however limited to the virtual neighborhood of the real node that is currently being deduplicated, and is cleared when we move on to the next real node.

\topic{Complexity} The runtime complexity is $O(n_r*d^4)$.

\topicul{Greedy Real Nodes First Algorithm}
In this less naive but still greedy approach, we consider each real node in sequence, and deduplicate it individually. Figure
    \ref{fig:real_nodes_first} shows an example, that we will use to illustrate the algorithm. The figure shows a real node $u_1$ that is connected
    to 5 virtual nodes, with significant duplication, and a deduplication of that node. Our goal here is to ensure that there are no duplicate edges
    involving $u_1$ -- we do not try to eliminate all duplication among all of $u_1$'s virtual nodes like in the naive approach. The core idea of
    this algorithm is that we consult a heuristic to decide whether to remove an edge to a virtual node and add the missing direct edges, or to keep the edge to the virtual node.

Let $\calV'$ denote the set of virtual nodes to which $u_s$ remains connected after deduplication,
and $\calV''$ denote the set of virtual nodes from which $u_s$ is disconnected; also, let $E$ denote the direct edges that we needed
to add from $u_s$ during this process. Our goal is to minimize the total number of edges in the resulting structure. This problem can be shown to be NP-Hard using a reduction from the \textit{exact set cover} problem.

We present a heuristic inspired by the standard greedy set cover heuristic which works as follows. We initialize $\calV' = \varnothing$, and $\calV''
= \calV$; we also logically add direct edges from $u_s$ to all its neighbors in $N(u_s)$, and thus $E = \{(u_s, x) | x \in \cup_{V \in \calV} O(V)\}$.
We then move virtual nodes from $\calV''$ to $\calV'$ one at a time. Specifically, for each virtual node $V \in \calV''$,  we consider moving
it to $\calV'$. Let $X = \cup O(\calV')$ denote the set of real nodes that $u$ is connected to through $\calV'$. In order to move
$V$ to $\calV'$, we must disconnect $V$ from all nodes in $O(V) \cap X$ -- otherwise there would be duplicate edges between $u$ and those nodes.
Then, for any $a, b \in O(V) \cap X$, we check if any other virtual node in $\calV''$ is connected to \textit{both} $a$ and $b$ -- if not, we must add the direct edge $(a,b)$. Finally, for $r_i \in O(V) - O(V) \cap X$, we remove all direct edges $(u, r_i)$.

The benefit of moving the virtual node $V$ from $\calV''$ to $\calV'$ is computed as the \textit{reduction} in the total number of edges in every scenario. We select
the virtual node with the highest benefit ($> 0$) to move to $\calV'$. If no virtual node in $\calV''$ has benefit $> 0$, we  move on to the next real node and leave $u$ connected to its neighbors through direct edges.


\topic{Complexity} The runtime complexity here is roughly $O(n_r*d^5)$.

\topicul{Greedy Virtual Nodes First Algorithm}
\label{subs:greedy_vfirst}
Exactly like the naive version above, this algorithm deduplicates the graph one virtual node at a time, maintaining a deduplicated partial graph at every step. We start with a graph containing only the real nodes and no virtual nodes, which is trivially deduplicated.
We then add the virtual nodes one at a time, always ensuring that the partial graph does not have any duplication. Let $V$ denote the
virtual node under consideration. Let $\calV = \{V_1, ..., V_n\}$ denote all the virtual nodes that share at least 2 real nodes with $V$
(i.e., $|O(V) \cap O(V_i)| \ge 2$). Let $C_i = O(V) \cap O(V_i)$, denote the real nodes to which both $V$ and $V_i$ are connected. At least $|C_i| -
1$ of those edges must be removed from $O(V)$ and $O(V_i)$ combined to ensure that there is no duplication.

The special case of this problem where $|C_i| = 2, \forall i$, can be shown to be equivalent to finding a {\em vertex cover} in a graph (we omit the
proof due to space constraints). We again adopt a heuristic inspired by the greedy approximation algorithm for vertex cover. Specifically, for each node
in $C_i$, we compute the $cost$ and the $benefit$ of removing it from any $O(V_i)$ versus from $O(V)$. The $cost$ of removing the node is computed as the number
of direct edges that need to be \textit{added} if we remove the edge to that virtual node, whereas the $benefit$ is computed as the \textit{reduction} in the total number of nodes in the intersection with $V_i$ ($\Sigma |C_i|$) (removing the node from $O(V_i)$ always
yields a $benefit$ of 1, whereas removing it from $O(V)$ may have a higher benefit). We then make a more informed decision and choose to remove an edge from a real node $rn$ that leads to the overall highest $benefit/ cost$ ratio.

\topic{Complexity} The runtime complexity here is: $O(n_vd(n_vd^2 + d))$.

\vspace{3pt}
\noindent
We note that, these complexity bounds listed here make worst-case assumptions and in practice, most algorithms run much faster.

\begin{figure}[t]
\begin{center}
\includegraphics[width=0.46\textwidth]{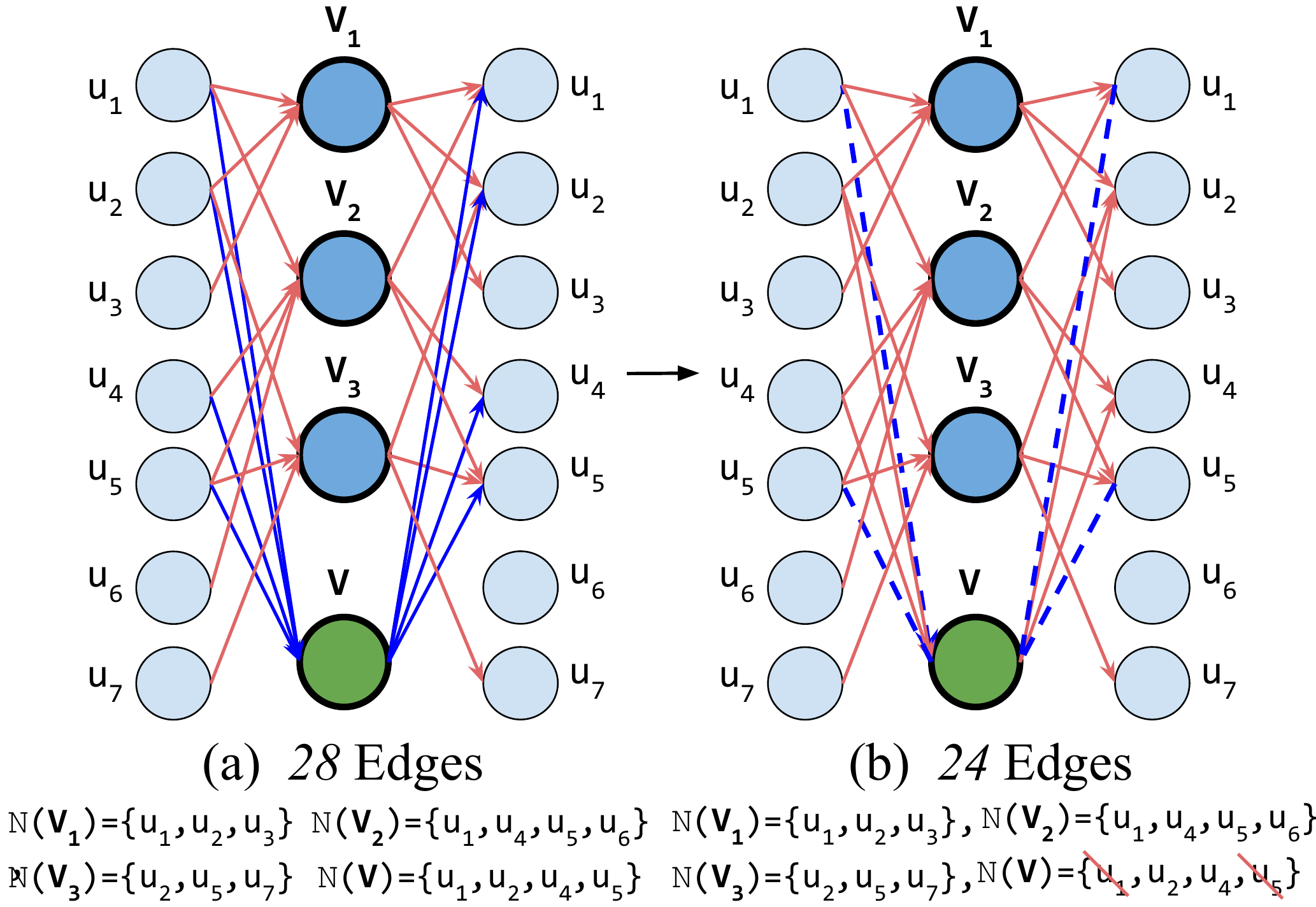}
\caption{Deduplication using Greedy Virtual Nodes First}
\vspace{-5pt}
\label{fig:virtual_nodes_first}
\end{center}
\end{figure}

\subsubsection{Multi-layer Condensed Graphs}
deduplicating multi-layer condensed graphs turns out to be significantly trickier and computationally more expensive than single-layer graphs. In
single layer graphs, identifying duplication is relatively straightforward; for two virtual nodes $V_1$ and $V_2$, if $O(V_1) \cap O(V_2) \ne \phi$ and $I(V_1) \cap I(V_2) \ne \phi$, then there is duplication. We keep the neighbor lists in sorted order, thus making these checks very fast. However, for multi-layer condensed graphs, we need to do expensive depth-first traversals to simplify identify duplication.

We can adapt the Naive Virtual Nodes First algorithm described above to the multi-layer case as follows. We (conceptually) add a dummy node $s$ to the condensed graph and add directed edges from $s$ to the $\__s$ copies of all the real nodes. We then traverse the graph in a depth-first fashion, and add the virtual nodes encountered to an initially empty graph one-by-one, while ensuring no duplication. However, this algorithm turned out to be infeasible to run even on small multi-layer graphs, and we do not report any experimental results for that algorithm.  Instead, we propose using either the BITMAP-2 approach for multi-layer graphs, or first converting it into a single-layer graph if possible (through expansion of all virtual nodes in all but one layer) and then using one of the algorithms developed above; note that the latter approach should only be considered if the expansion does not result in a space explosion.

\vspace{-5pt}
\section{Experimental Study}
\label{sec:experiments}
In this section, we provide a comprehensive experimental evaluation of \graphgen using several real-world and synthetic datasets.
We first present a detailed study using 4 small datasets. 
We then compare the performance of the different deduplication algorithms, and present an analysis using much larger
datasets, but for a smaller set of representations.
All the experiments were run on a single machine with 24 cores running at 2.20GHz, and with 64GB RAM.

\subsection{Small Datasets}
\label{sub:smalldatasets}
First we present a detailed study using 4 relatively-small datasets.
We use representative samples of the \textit{DBLP} and {\em IMDB} datasets in our study (Table \ref{tab:datasets}), extracting {\em co-author} and {\em co-actor} graphs respectively. We also generated a series of synthetic graphs so that we can better understand the differences between
the representations and algorithms on a wide range of possible datasets, with varying numbers of real nodes and virtual nodes, and varying degree distributions and densities.
Since we need the graphs in a condensed representation, we cannot use any of the existing random graph generators for this purpose. Instead, we built a synthetic graph generator,
which we sketch in Appendix~\ref{app:smalldatasets}. 

{
\begin{table}
\center
    \small
\csvreader[tabular=|c|c|c|c|c|,
    table head=\hline \textbf{Dataset} & \textbf{Real Nodes} & \textbf{Virt Nodes} & \textbf{Avg Size} & \textbf{EXP Edges}\\\hline,
    late after line=\\\hline]%
{numbers/datasets.csv}{Dataset=\dataset,RealNodes=\realNodes,VirtalNodes=\virtNodes,AvgCliqueSize=\avgcs,ExpandedEdges=\exp}%
{\dataset & \num[group-separator={,}]{\realNodes} & \num[group-separator={,}]{\virtNodes} & $\avgcs$ & \num[group-separator={,}]{\exp}}
\caption{Small Datasets: \textbf{avg size} refers to the average number of real nodes contained in a virtual node}
\label{tab:datasets}
\end{table}
}

\subsubsection{Compression Performance}
\label{sec:vminer}
We begin with comparing the graph sizes for the different representations for each dataset. In addition to the in-memory representations presented
in this paper, we also implemented and compared against a prior graph compression algorithm, called {\em VMiner} ({\em Virtual Node Miner})~\cite{buehrer2008scalable}. VMiner
uses {\em frequent pattern mining} to identify {\em bi-cliques} in the graph, i.e., groups of nodes $A$ and $B$, such that for $u \in A, v \in B$, there
is an edge $u \rightarrow v$. It then repeatedly replaces such bicliques with {\em virtual nodes}; i.e., it adds a new virtual node $C$ to the graph, adds an edge
$u \rightarrow C, \forall u \in A$ and $C \rightarrow v, \forall v \in B$, and deletes all edges from $A$ to $B$.
It makes multiple passes through the graph, iteratively reducing its size.
The final representation thus looks very similar to DEDUP-1, and is also duplication-free. VMiner has several parameters which we exhaustively tried out
combinations of, for each of our datasets, (following the guidance in the paper) and picked the best. \ul{Note that using VMiner requires us to first
expand the graph, which makes it infeasible for several of the large datasets discussed in Section 6.2.}

\begin{figure}[t]
\begin{center}
\includegraphics[width=0.42\textwidth]{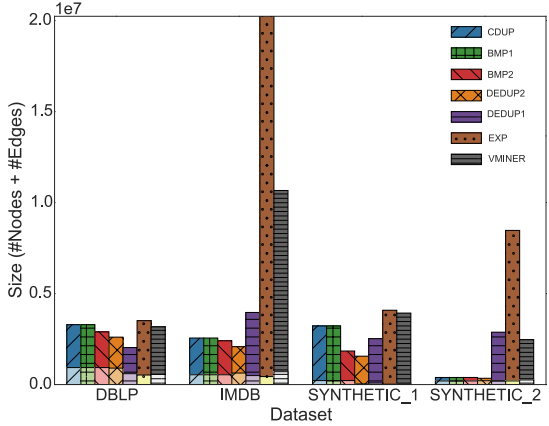}
\caption{Comparing the in-memory graph sizes for different datasets; the bottom (lighter) bars show the number of nodes.}
\vspace{-5pt}
\label{fig:sizes}
\end{center}
\end{figure}

Figure \ref{fig:sizes} shows how the different algorithms fare against each other. For each algorithm and each dataset, we report the total number of nodes and edges, and also show the breakdown between them; the algorithm used for \textit{DEDUP-1} was Greedy Virtual Nodes First, described in Section~\ref{subs:greedy_vfirst}.
When the average degree of virtual nodes is small and there is a large number of virtual nodes (as is the case with \textit{DBLP} and \textit{Synthetic\_1}), we observe that there is a relatively small difference in the size of the condensed and expanded graphs, and deduplication (\textit{DEDUP-1} and \textit{DEDUP-2}) actually results in an even smaller footprint graph.

On the other hand, the \textit{IMDB} dataset shows a 8-fold difference in size between EXP and C-DUP and over a 5-fold difference with all other
representations. 
\textit{Synthetic\_2} portrays the amount of
compression possible in graphs with very large, overlapping cliques. The BITMAP representations prevail here as well; however this dataset also shows how the \textit{DEDUP-2} representation can be significantly more compact than \textit{DEDUP-1}, while maintaining its natural, more portable structure compared to the BITMAP representations.
\ul{As we can see, VMiner not only requires expanding the graph first, but also generally finds a much worse representation than DEDUP-1. This corroborates our hypothesis that
working directly with the implicit representation of the graph results in better compression.}

We also measured actual memory footprints for the same datasets, which largely track the relative performance shown here, with one major difference being
that BITMAP representations perform a little worse because of the extra space required for storing the bitmaps. 
We report memory footprints for larger datasets in Section \ref{sec:large}.

\begin{figure}[b]
    \centering
    \begin{subfigure}[t]{0.23\textwidth}
        \includegraphics[width=\textwidth]{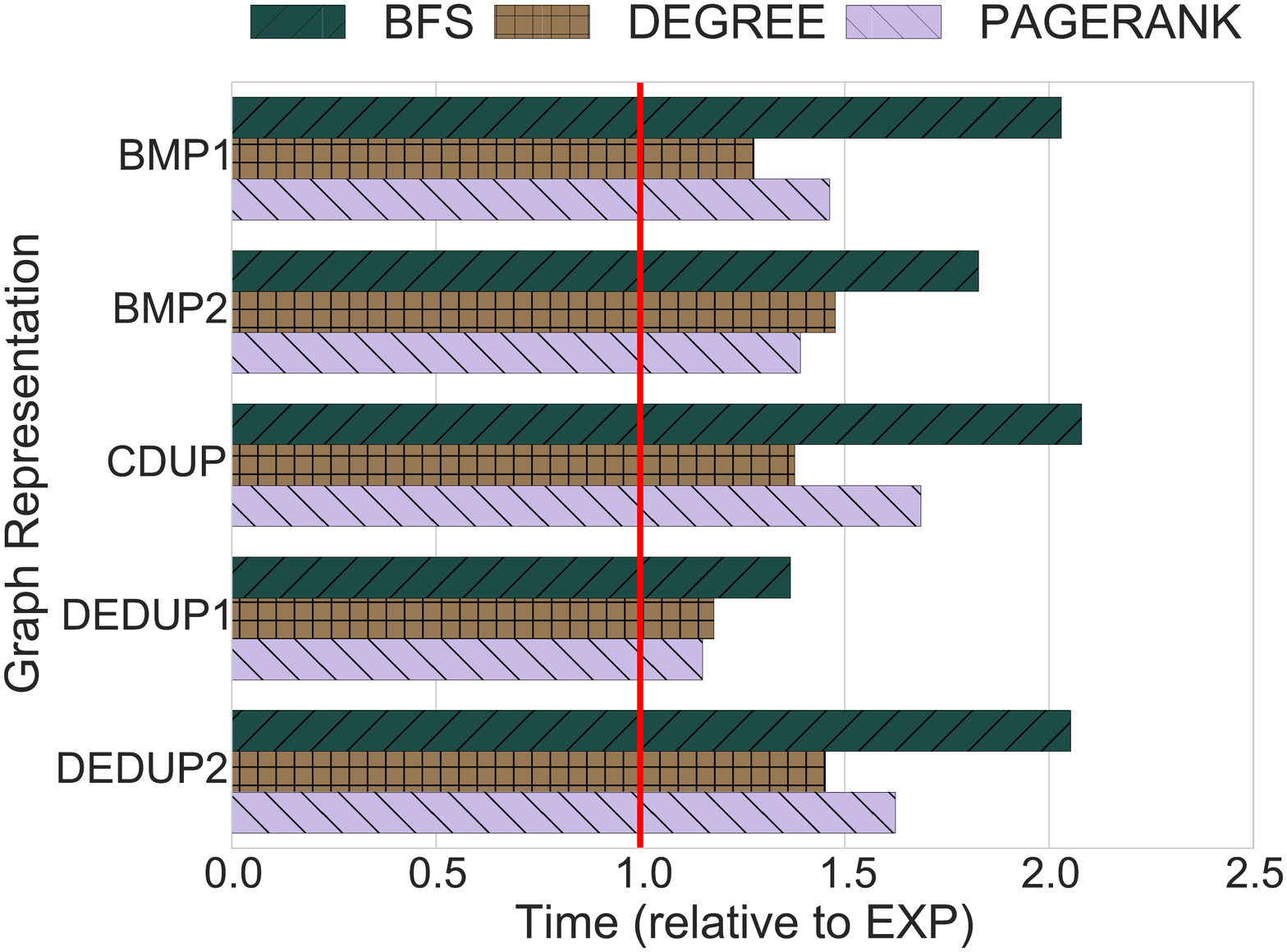}
        \caption{DBLP}
    \end{subfigure}
    \begin{subfigure}[t]{0.23\textwidth}
        \includegraphics[width=\textwidth]{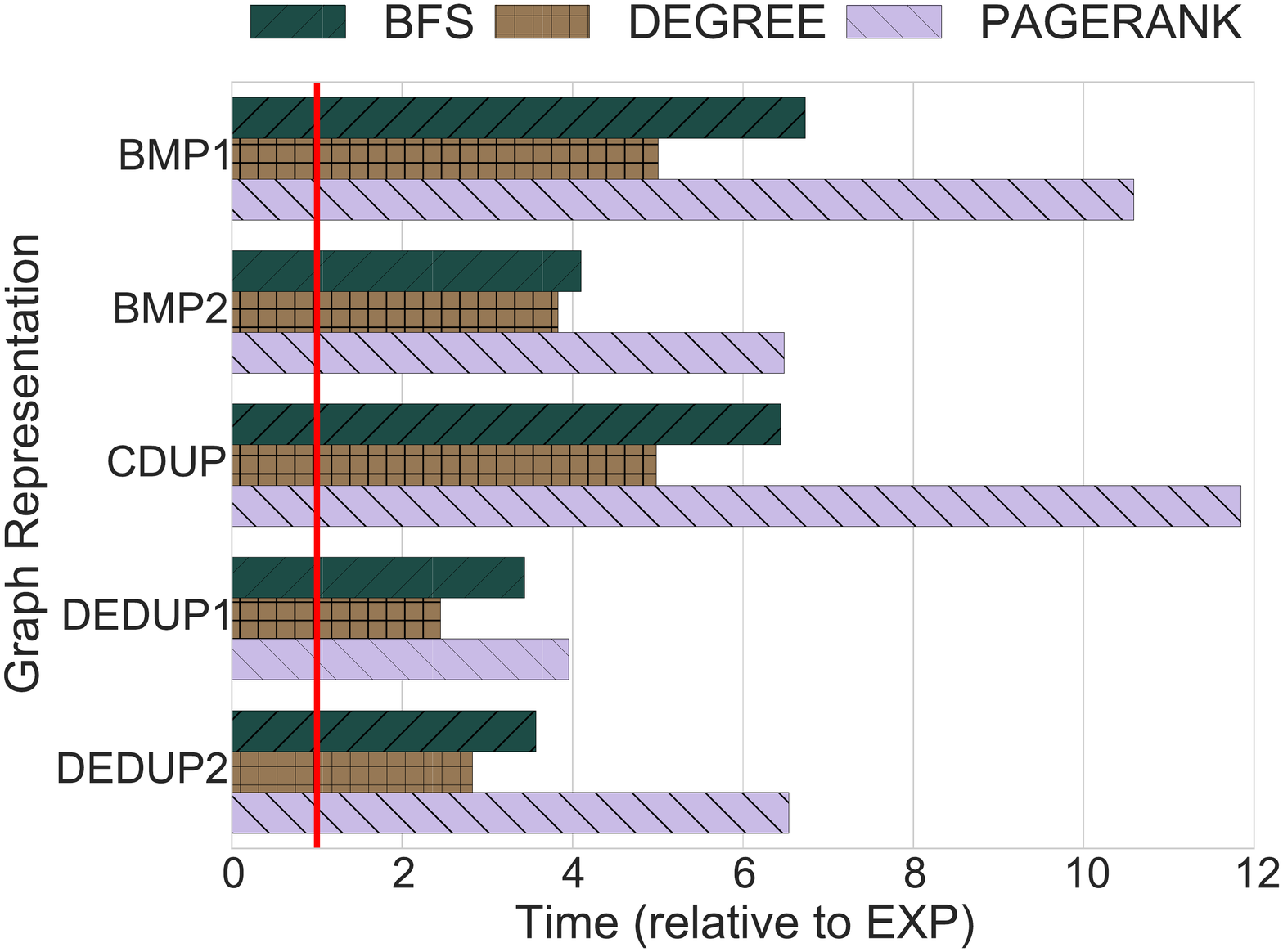}
        \caption{Synthetic\_1}
    \end{subfigure}
    \caption{Performance of Graph Algorithms on Each Representation for two datasets (vertical red line represents EXP)}\label{fig:algorithms}
     \vspace{-5pt}
\end{figure}

\subsubsection{Graph Algorithms Performance}
Figure \ref{fig:algorithms} shows the results of running 3 different graph algorithms on the different in-memory representations. We compared the performance of \textit{Degree} calculation, \textit{Breadth First Search (BFS)} starting from a single node, as well as \textit{PageRank} on the entire graph. Again, the results shown are normalized to the values for the full EXP representation.
Degree and PageRank were implemented and run on our custom vertex-centric framework described in Section~\ref{subs:analyzing}, while BFS was run in a single threaded manner starting from a single random node in the graph, using our Graph API\cutforrevision{to operate directly on top of each of the representations}. Again,
       the BFS results are the mean of runs on a specific set of 50 randomly selected real nodes on all of the representations, while the PageRank are an average of 10 runs.

       We also ran a comprehensive set of microbenchmarks comparing the performance of the basic graph operations against the different representations. Those results can be found in Appendix \ref{sec:microbenchmarks}, and as can be seen there, BFS and PageRank both follow the trends of the micro-benchmarks in terms of differences 
       between representations.

For \textit{IMDB} and \textit{Synthetic\_2}, both of which yield very large expanded graphs, we observed little to no overhead in real world performance compared to
EXP when actually running algorithms on top of these representations, especially when it comes to the BITMAP and DEDUP-1 representations (we omit these graphs). DBLP and Synthetic\_1 datasets portray a large gap in performance compared to EXP; this is because these datasets consist of a large number of small virtual nodes, thus increasing the average number of virtual nodes that need to be iterated over for a single calculation. This is also the reason why DEDUP-1 and BITMAP-2 typically perform better; they feature a smaller number of virtual neighbors per real node than representations like C-DUP and BITMAP-1, and sometimes DEDUP-2 as well.

%

{
\begin{table*}[t]
\centering
\small
\begin{tabu} to \textwidth{|l|l|l|l|l|l|l|l|l|l|l|l|l|l|}
    \hline
    \multirow{2}{*}{\bf Dataset} &
      \multicolumn{4}{c|}{\ul{\bf CDUP}} &
      \multicolumn{5}{c|}{\ul{\bf BMP-DEDUP}} &
      \multicolumn{4}{c|}{\ul{\bf EXP}} \\
    & Degree & PR & BFS & \textbf{Mem (GB)} & Degree & PR & BFS & \textbf{Mem (GB)} & Dedup Time & Degree & PR & BFS & \textbf{Mem (GB)} \\
    \hline
    {\bf Layered\_1} & 40 & 1211 & 382 & \textbf{1.421} & 30 & 1025 & 284 & \textbf{2.737} & 1714 & DNF & DNF & DNF & \textbf{>64} \\
    \hline
    {\bf Layered\_2} & 12 & 397 & 129 & \textbf{1.613} & 10 & 339 & 111 & \textbf{2.258} & 553 & 11 & 83 & 85 & \textbf{19.798} \\
    \hline
    {\bf Single\_1} & 2 & 30 & 0.01 & \textbf{1.276} & 1.8 & 25 & 0.02 & \textbf{1.493} & 10.4 & 1.6 & 14.7 & 0.01 & \textbf{1.2} \\
    \hline
    {\bf Single\_2} & 202 & DNF & 1.3 & \textbf{9.901} & 81 & 3682 & .12 & \textbf{13.042} & 5871 & DNF & DNF & DNF & \textbf{>65} \\
    \hline
    {\bf TPCH} & 3.5 & 58 & 86 & \textbf{.023} & 0.4 & 6 & 8.6 & \textbf{.049} & 1207 & 1.470 & 8 & 16 & \textbf{7.398} \\
    \hline
  \end{tabu}
\vspace{5pt}
  \caption{Comparing the performance (running times in seconds, and memory consumption in GB) of C-DUP, BITMAP, and EXP on large datasets; the table also shows the time required for bitmap de-duplication (DNF $\rightarrow$ {\em did not finish} in reasonable time).} 
  \label{tab:largeGraphs}
  \vspace{-10pt}
\end{table*}
}


\begin{figure}[h]
    \centering
    ~ 
      \hspace{-20pt}
    \begin{subfigure}[t]{0.2375\textwidth}
        \includegraphics[width=\textwidth]{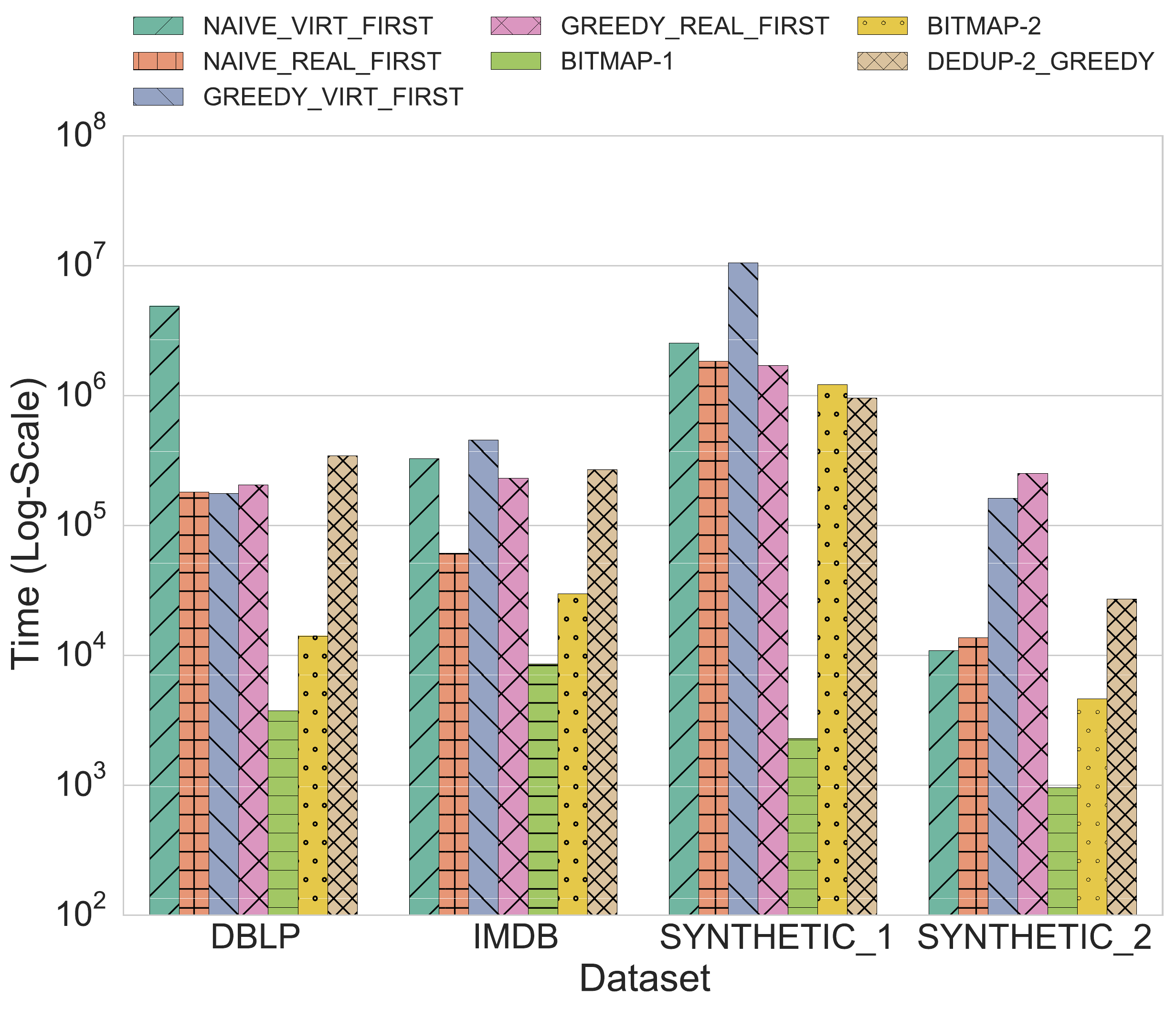}
        \caption{}
        \label{fig:dedup_time}
    \end{subfigure}
    ~ 
    \begin{subfigure}[t]{0.2375\textwidth}
        \includegraphics[width=\textwidth]{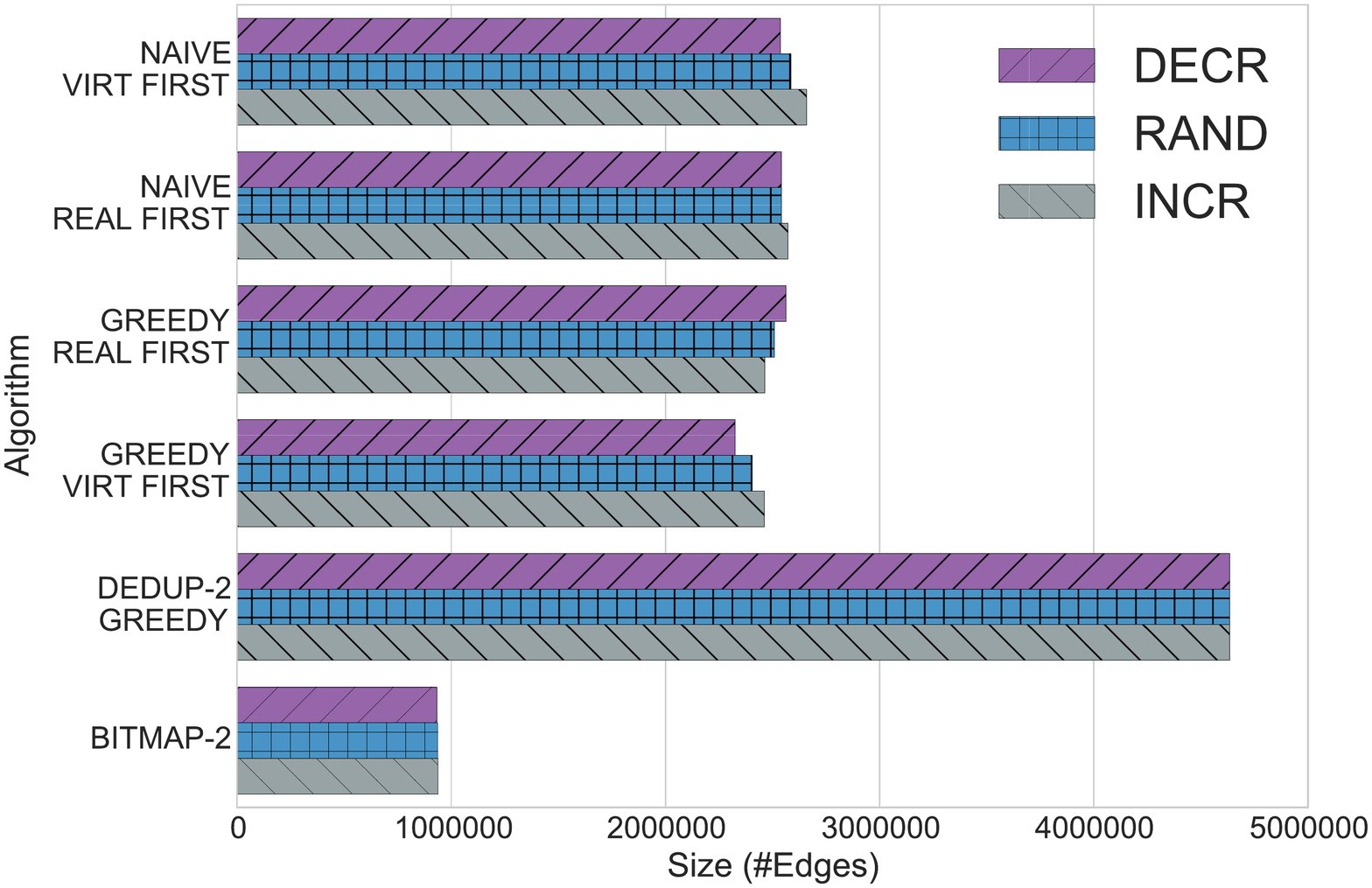}
        \caption{}
        \label{fig:dedup_orderings}
      \end{subfigure}
    \caption{Deduplication Performance Results (a) Deduplication time comparison between algorithms. Random (RAND) vertex ordering was used where applicable, (b) Small variations caused by node ordering in deduplication}
    \vspace{-5pt}
\end{figure}

\subsubsection{Comparing Deduplication Algorithms}
\label{sub:algoriths_discussion}

Figure \ref{fig:dedup_time} shows the running times for the different deduplication algorithms (on a log-scale). As expected, BITMAP-1 is the fastest of the algorithms, whereas the DEDUP-1 and DEDUP-2 algorithms take significantly more time. We note however that deduplication is a one-time cost, and the overhead of doing
so may be acceptable in many cases, especially if the extracted graph is serialized and repeatedly analyzed over a period of time. Finally, Figure \ref{fig:dedup_orderings} shows how the performance of the various algorithms varies depending on the processing order. We did not observe any noticeable
differences or patterns in this performance across various datasets, and recommend using the random ordering for robustness.

\subsection{Large Datasets}
\label{sec:large}
To reason about the practicality and scalability of \graphgen, we evaluated its performance on a series of datasets
that yielded larger and denser graphs (Table \ref{tab:largeGraphs}). Datasets \textit{Layered\_1} and \textit{Layered\_2} are synthetically
generated multi-layer condensed graphs, while \textit{Single\_1}, \textit{Single\_2} are standard single-layer condensed graphs (cf. Appendix \ref{app:largedatasets} for details on
how these datasets were generated). At this scale, only the C-DUP, BITMAP-2, and EXP are typically feasible options, since none of the deduplication algorithms (targeting
DEDUP-1 or DEDUP-2) run in a reasonable time.

Comparing the memory consumption, we can see that we were not able to expand the graph in 2 of the cases, since it consumed more memory than available
($> 64GB$); in the remaining cases, we see that EXP consumes more than 1 or 2 orders of magnitude more memory. In one case, EXP was actually smaller
than C-DUP; our preprocessing phase (Section~\ref{sec:extraction}), which was not used for these experiments, would typically expand the graph in such cases.
Runtimes of the graph algorithms show the patterns we expect, with EXP typically performing the best (if feasible), and BITMAP somewhere in between EXP and C-DUP (in some
cases, with an order of magnitude improvement). Note that: we only show the base memory consumption for C-DUP -- the memory consumption can be significantly higher
when executing a graph algorithm because of on-the-fly deduplication that we need to perform. In particular,
C-DUP was not able to complete PageRank for Single\_2, running out of memory. 

As these experiments show, datasets don't necessarily have to be large in order to hide some very dense graphs, which would normally be extremely expensive to extract and analyze. This is shown in the TPCH dataset where we extracted a graph of customers who have bought the same item. With \graphgen, we are able to load them into memory and with a small deduplication cost, are able to achieve comparable iteration performance that allows users to explore, and analyze them in a fraction of the time, and using a fraction of the machine's memory that would be initially required.

\begin{figure*}[t]
    \centering
    \begin{subfigure}[t]{0.24\textwidth}
        \includegraphics[width=\textwidth]{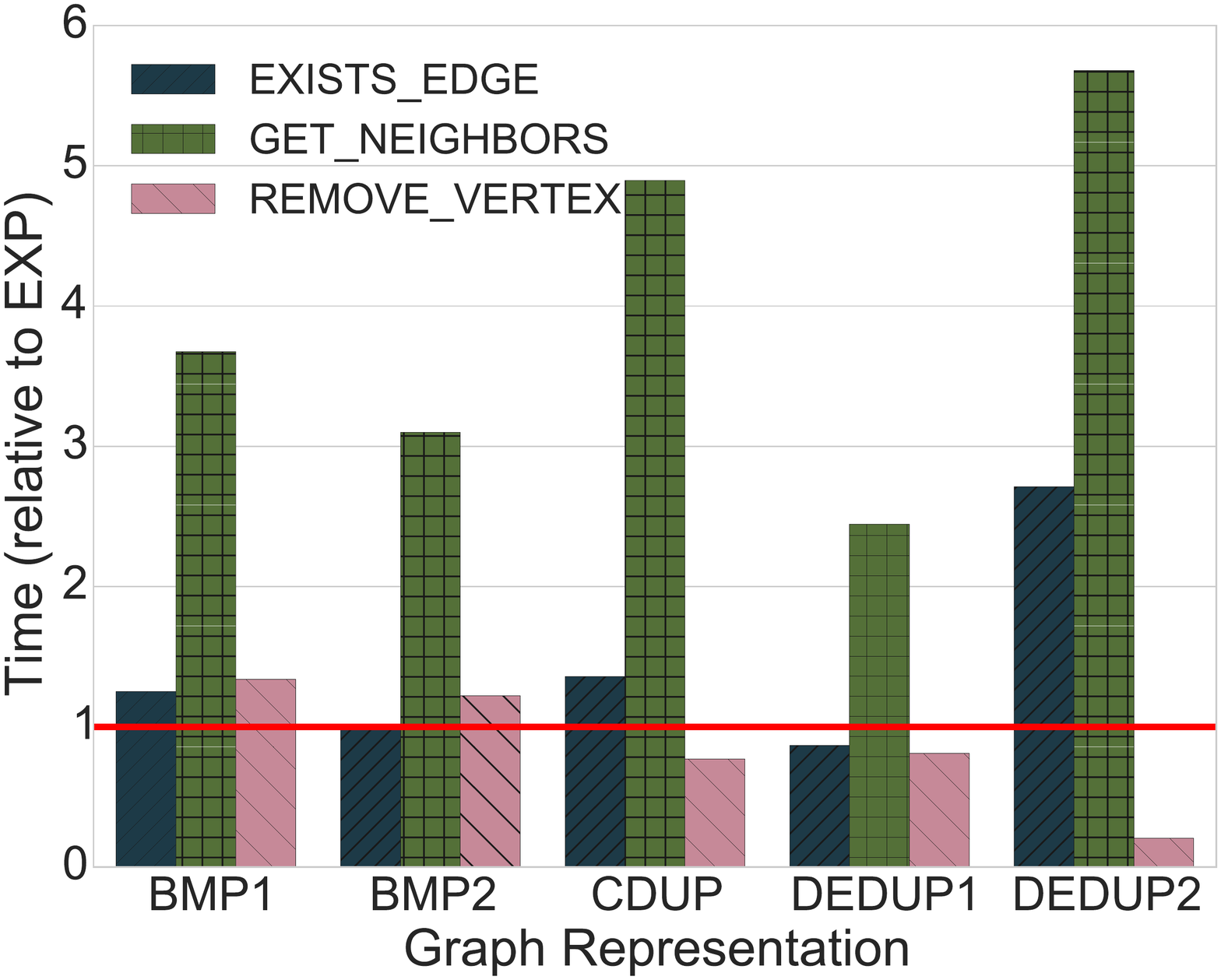}
        \caption{DBLP}
    \end{subfigure}
    \begin{subfigure}[t]{0.24\textwidth}
        \includegraphics[width=\textwidth]{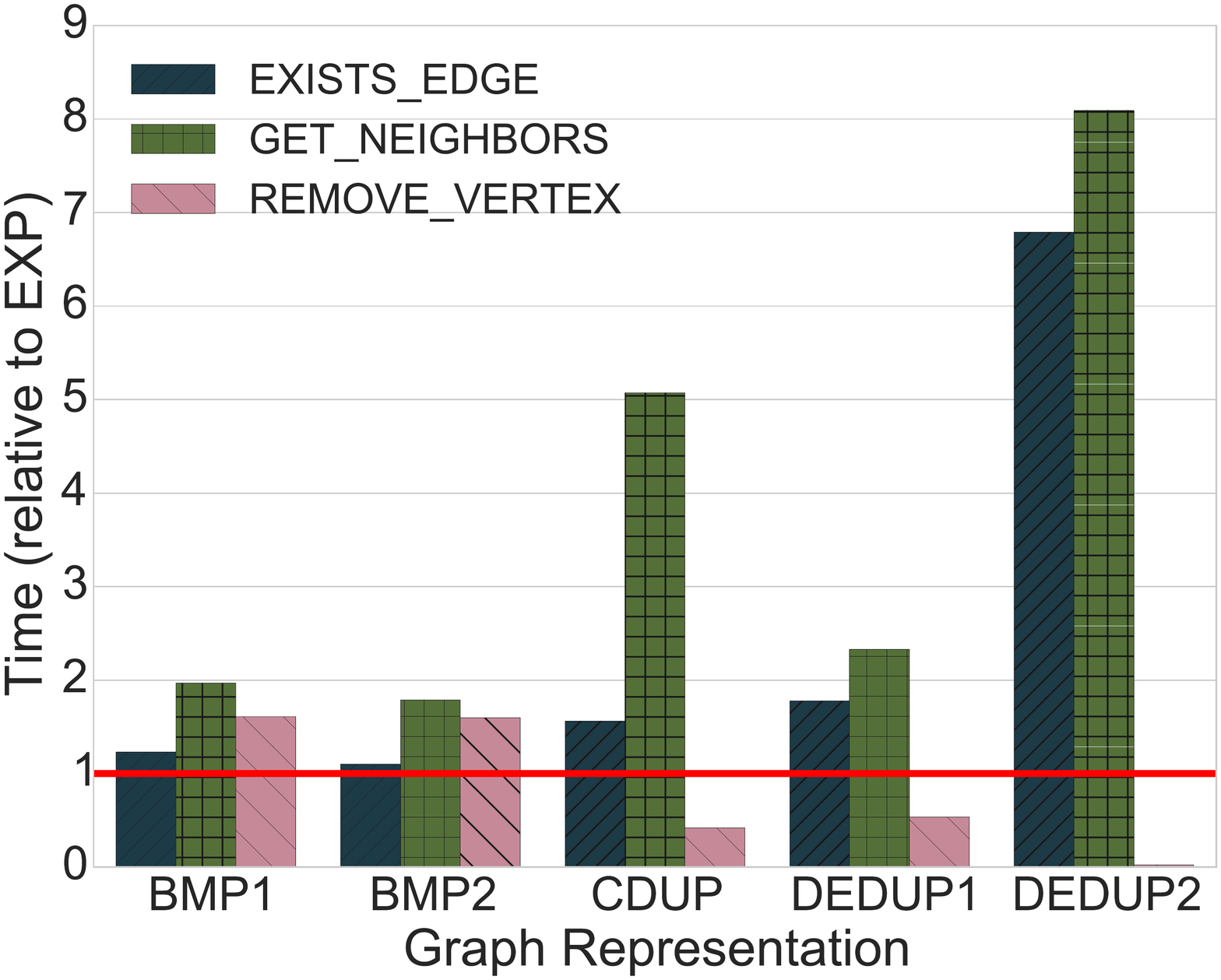}
        \caption{IMDB}
    \end{subfigure}
    \begin{subfigure}[t]{0.24\textwidth}
        \includegraphics[width=\textwidth]{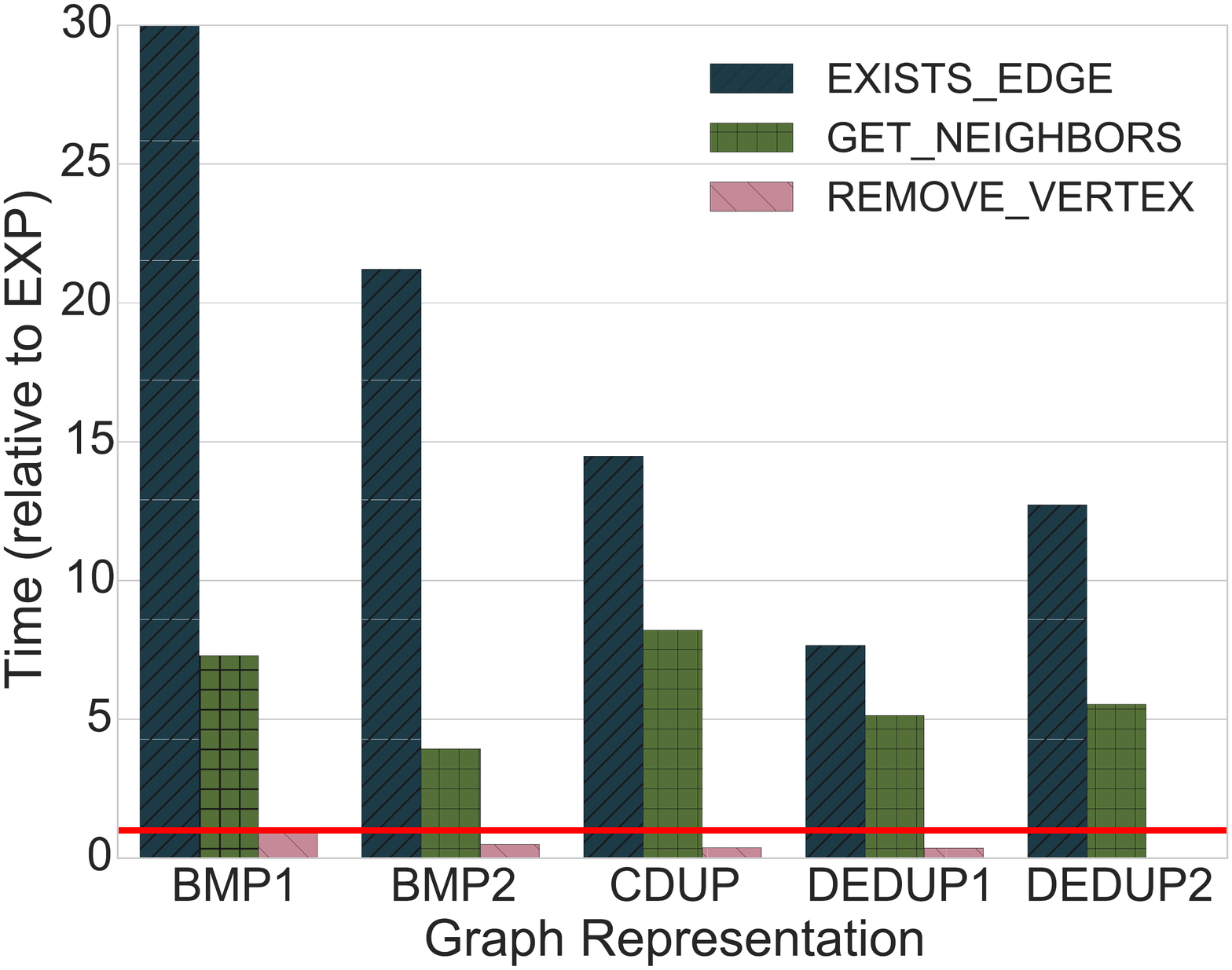}
        \caption{Synthetic\_1}
    \end{subfigure}
    \begin{subfigure}[t]{0.24\textwidth}
        \includegraphics[width=\textwidth]{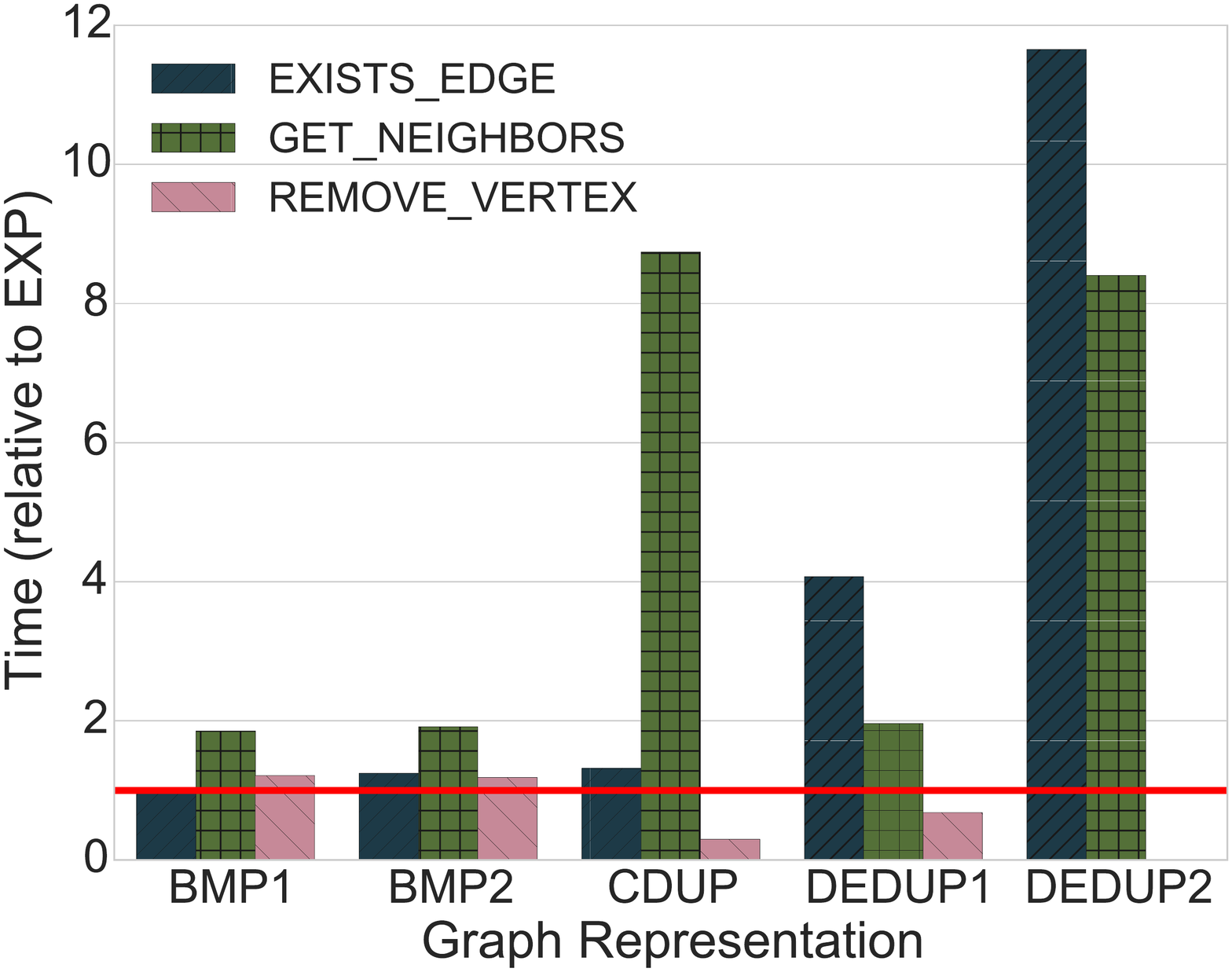}
        \caption{Synthetic\_2}
    \end{subfigure}
    \caption{Microbenchmarks for each representation}\label{fig:microbenchmarks}
\end{figure*}

\subsection{Microbenchmarks}
\label{sec:microbenchmarks}

We conducted a complete set of micro-benchmarks on each function in our Graph API described in Section~\ref{sec:inmemory}. Figure \ref{fig:microbenchmarks} shows the results for some of the more interesting graph operations. The results shown are normalized using the values for the full \textit{EXP}
representation, which typically is the fastest and is used as the baseline. Since most of these operations take micro-seconds to complete, to ensure validity in the results, the metrics shown are the result of the mean of $3000$ repetitions for each operation, on a specific set of the same $3000$ randomly selected nodes for each dataset.

Iteration through each real node's neighbors via the \textsc{getNeighbors()} method is naturally more expensive on all other representations compared
to the expanded graph. This portrays the natural tradeoff of extraction latency and memory footprint versus performance that is offered by these
representations. \textit{DEDUP-2} is usually least performant here because of the extra layer of indirection that this representation introduces. 
\textit{DEDUP-1} is typically more performant than the \textit{BITMAP} representations in datasets where there is a large number of small cliques.

In terms of the \textsc{existsEdge()} operation, we have included auxiliary indices in both virtual and real vertices, which allow for rapid checks on whether a logical edge exists between two real nodes. Latency in this operation is relative to the total number of virtual nodes, the indexes of which need to be checked.
The \textsc{removeVertex()} operation is actually \textit{more efficient} on the C-DUP, DEDUP-1 and DEDUP-2 representations than EXP. In order for a vertex to be removed from the graph, explicit removal of all of its edges is required. In representations like DEDUP-1 and DEDUP-2, that employ virtual nodes, we need to remove a smaller number of edges on average in the removal process. DEDUP-2 is most interesting here because a real node is always connected to only 1 virtual node, therefore the removal cost is constant .

\subsection{Integration with Giraph} 
\label{sec:giraph}

The wide array of representations we propose are significantly more memory-efficient than storing the entire graph (i.e., EXP representation). Further, the
C-DUP representation is the easiest and fastest to obtain from the relational database
(Table \ref{tab:benefits}) and is usually the most memory-efficient.
\cutforrevision{
Past work on graph compression aims to compress the expanded graph into one with a smaller
memory footprint. These techniques require the initial storage of the entire
expanded graph, which is then processed and compressed, potentially requiring
memory even larger than the expanded graph itself. In our setting, this would
defeat the purpose of the system as our system aims at efficiently extracting
and operating on top of graphs in relational datasets without requiring
extraction and storage of the expanded graph.
}
Our representations are quite generic and can be ported 
to other graph processing systems with varying degrees of difficulty. In Table~\ref{tab:giraphExperiments}, we
showcase the results of running {\em Degree, PageRank} and {\em Connected Components} on three representations
on a prototype port to Apache Giraph, for a diverse set of synthetic datasets
created using our generator (cf. Appendix~\ref{app:smalldatasets}). The datasets \textit{S1-2} were created by maintaining the number of real and
virtual nodes static, and incrementally increasing the average
\textit{size} of each virtual node. For datasets \textit{N1-2} we
maintained the average size of each virtual node static, but increased the
number of real and virtual nodes. We also ran these experiments over the co-actor graph extracted from portion of the \textit{IMDB} (Table~\ref{tab:benefits}).

We implemented these three algorithms as efficiently as possible for each representation. As seen in Table~\ref{tab:giraphExperiments}, our BITMAP representation almost always outperforms EXP and DEDUP-1, while requiring up to an order of magnitude less memory. Specifically, when using BITMAP, \textit{Connected Components} received a speedup due to the fact that it is \textit{duplicate-insensitive} and can be run directly over C-DUP. Running \textit{PageRank} and \textit{Degree} requires a deduplicated graph, and further, correct execution over DEDUP-1 and BITMAP  requires \textit{twice} the number of supersteps.
By implementing
 \textit{message aggregation} at each virtual node, we were able to decrease the
 number of messages that need to be passed per superstep to only $2 * \#edges$,
 which resulted in a \textit{speedup} over EXP for the larger datasets. Even though DEDUP-1 was not able to achieve a significant compression over EXP for these datasets, it still outperformed EXP on the larger datasets.

It's interesting to observe the different performance trends seen in the \textit{IMDB} graph. Here we can see that DEDUP-1 ends up being the best alternative in terms of both memory consumption and running times while BITMAP often ends up being second or third in comparison. This skew in the results comes from the difference between the number of nodes in these datasets. The BITMAP representation ends up having nearly \textit{twice} the number of nodes than EXP, and also shows a substantial difference from DEDUP-1; this difference is due to the \textit{virtual nodes} we need to store for both DEDUP-1 and BITMAP. Also, the fact that we see memory consumption for BITMAP being close to or surpassing EXP in many situations is due to the fact that storing more \textit{nodes} (and in the case of BITMAP, several bitmaps associated with each node), incurs a larger overhead than storing more \textit{edges}. Also, in comparison with the other datasets, the difference in the number of edges between the representations is significantly smaller, which also plays into the fact that BITMAP does not provide as big of an advantage as for the previous datasets. Nevertheless, in situations where a lot of messages need to be sent and received (like when running \textit{PageRank}), the benefits of BITMAP already start showing up, with BITMAP being on par with DEDUP-1, and better than EXP.

One of the fundamental issues that have to be dealt with regarding
ports like this is the fact that, some vertex centric algorithms assume
direct access to a node's immediate neighbors and therefore
assume that calculating the degree for each node is trivial and fast.
In the case of our representations, while calculating the degree is
trivial, there is no direct access to the immediate neighbors at each
node, and therefore the degree cannot be computed on the fly when
the vertex-centric framework is used. When the degree needs to
be used continuously in a vertex-centric program (e.g., PageRank),
it needs to be pre-computed and stored as a vertex property once
before the computation begins, otherwise we would need an entire
superstep every time only to compute the degree before continuing
with the next iteration of the program.

\begin{table}
\center
    \small
\begin{tabular}{|c|c|c|c|c|c|c|c|}
\hline
\textbf{Data} & \textbf{Repr} & \multicolumn{2}{c|}{\textbf{Degree}} & \multicolumn{2}{c|}{\textbf{ConComp}} & \multicolumn{2}{c|}{\textbf{PageRank}} \\
\textbf{Set} &  & \emph{time} & \emph{mem}  & \emph{time} & \emph{mem} & \emph{time} & \emph{mem} \\
\hline
S1&EXP& 61 & 237& 90 & 202& \textbf{245} & 526\\
&DEDUP1& 54 & 227& \textbf{81} & 171& 311 & 484\\
&BMP&\textbf{50} & \textbf{134}&82 & \textbf{72}&256 & \textbf{156}\\
\hline
S2&EXP& \textbf{294} & 2,879& 498 & 2,869& 3,287 & 9,164\\
&DEDUP1& 335 & 2,582& 460 & 2,573& 3,049 & 8,126\\
&BMP&311 & \textbf{186}& \textbf{335} & \textbf{163}&\textbf{812} & \textbf{293}\\
\hline
N1&EXP& 142 & 1,109& 241 & 1,088& 1,456 & 3,389\\
&DEDUP1& 141 & 926& 483 & 901& 1,317 & 2,874\\
&BMP&\textbf{131} & \textbf{219}&\textbf{149} & \textbf{150}&\textbf{469} & \textbf{377}\\
\hline
N2&EXP& 268 & 2,710& 593 & 2,690& 4,493 & 8,432\\
&DEDUP1& 312 & 2,216& 495 & 2,194& 3,726 & 6,892\\
&BMP&\textbf{257} & \textbf{479}&\textbf{280} & \textbf{347}& \textbf{824} & \textbf{691}\\
\hline
IMDB & EXP & \textbf{78} & 586 & \textbf{193} & 749 & 861 & 1178\\
& DEDUP1 & 85 & \textbf{553} & 194 & \textbf{594} & \textbf{802} & \textbf{764}\\
& BMP & 146 & 952 & 291 & 1038 & 807 & 1185\\
\hline
\end{tabular}
\caption{Experiments on Giraph showing the running \textit{time(s) / memory(MB)} for different representations and algorithms}
\label{tab:giraphExperiments}
\vspace{-12pt}
\end{table}

\begin{table}
\center
    \small
\csvreader[tabular=|c|c|c|c|c|,
    table head=\hline \textbf{Dataset} & \textbf{Repr} & \textbf{All Nodes} & \textbf{Virt Nodes} & \textbf{Edges} \\\hline,
    late after line=\\\hline]%
{numbers/giraph_datasets.csv}{dataset=\dataset,repr=\repr,realnodes=\realnodes,virtualnodes=\virtualnodes,edges=\edges}
{\dataset & \repr & \num[group-separator={,}]{\realnodes} & \num[group-separator={,}]{\virtualnodes}  & \num[group-separator={,}]{\edges}}
\caption{Descriptions of the datasets used for experiments with Giraph}
\label{tab:giraphDatasets}
 \vspace{-12pt}
\end{table}

\subsection{Discussion}

Our experimental evaluation illustrates the pros and cons of the different representations, which leaves us with the question of which representation
to choose in a particular setting. Note that, in several of the experiments, we did not use the preprocessing step (Section 4.2) to allow us to
more properly compare the different representations. In practice, however, our system always uses the preprocessing step, and we further suggest
that the graph be expanded if the memory increase is not substantial, e.g., less than 20\% (the size of the expanded graph can be calculated relatively
quickly from the C-DUP representation). If expanding the graph is not an option, then the system needs to choose
between C-DUP, BITMAP-2, DEDUP-1, DEDUP-2. These representations are better in different settings, and thus the choice comes down to the use-case. For
graph algorithms that don't touch a large portion of the graph, C-DUP is the best option (e.g., {\em breadth-first search}). BITMAP-2 is preferred for
more complex graph algorithms that might make multiple passes on the graph (e.g., {\em PageRank}). On the other hand, DEDUP-1 and DEDUP-2 should be
used if multiple graph algorithms need to be run over a period of time, to amortize the cost of constructing those; in those cases, it might even
be a good idea to store the deduplicated graphs back into the database, with the caveat that changes to the underlying relations would require
updating the graph. Our system allows making these choices easily and on a per-algorithm basis.

\vspace{-5pt}
\section{Conclusion}
In this paper, we presented \graphgen, a system that enables users to analyze the implicit
interconnection structures between entities in normalized relational databases, without the need to
extract the graph structure and load it into specialized graph engines. \graphgen can interoperate
with a variety of graph analysis libraries and supports a standard graph API, breaking down the
barriers to employing graph analytics. However, these implicitly defined graphs can often be orders
of magnitude larger than the original relational datasets, and it is often infeasible to extract
or operate upon them. We presented a series of in-memory condensed representations and deduplication
algorithms to mitigate this problem, and showed how we can efficiently run graph algorithms on such graphs
while requiring much smaller amounts of memory. The choice of which representation to use depends on
the specific application scenario, and can be made at a per dataset or per analysis level. The
deduplication algorithms that we have developed are of independent interest, since they result in
a compressed representation of the extracted graph. Some of the directions for future work include
tuning the selectivity estimates for complex extraction queries, and extending our deduplication
algorithms to handling general directed, heterogeneous graphs.

\topicul{Acknowledgments} This work was supported by NSF under grant IIS-1319432, and an IBM Collaborative Research Award. We thank the anonymous
referees for valuable suggestions that helped us improve the paper substantially.

{

}


\appendix

\section{Connection with Factorized Representations}
\label{app:factorized}
\begin{figure}[h]
\begin{center}
\includegraphics[width=0.45\textwidth]{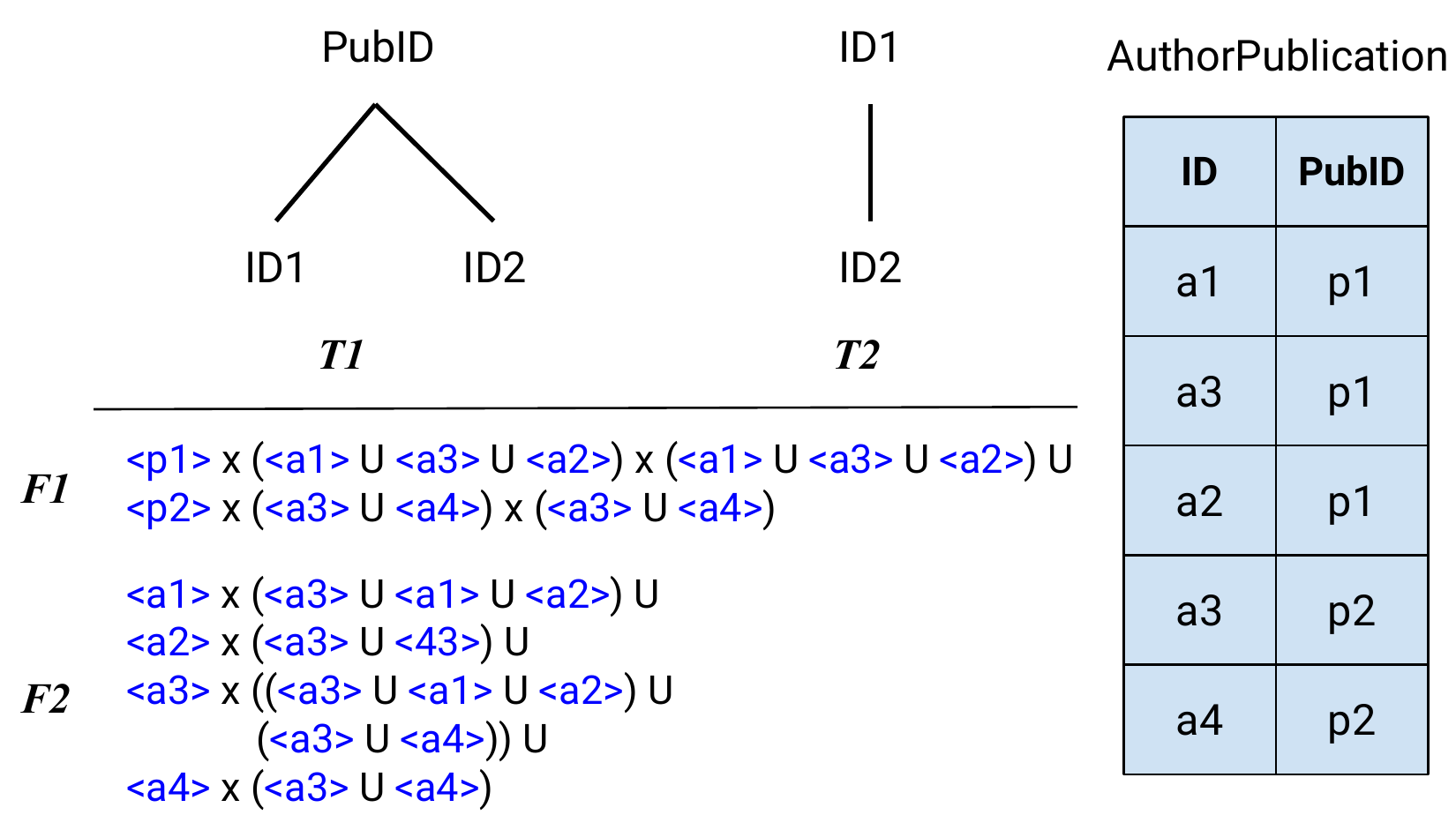}
\caption{\textit{T1} results in factorization \textit{F1} (equivalent to C-DUP). \textit{T2} results in factorization \textit{F2} which is equivalent to the (expanded) graph}
\label{fig:ftree}
\end{center}
\end{figure}

We further elaborate on the connection with the recent work on factorized representations of query results~\cite{olteanu2015size}, using the {\em co-authors} graph extraction query.
\begin{lstlisting}[breaklines,basicstyle=\ttfamily]
   Edges(ID1,ID2) :- AuthorPublication(ID1,PubID), AuthorPublication(ID2,PubID).
\end{lstlisting}
The final query result here (i.e., the list of edges) is equivalent to the expanded graph, and our techniques in this paper are focused on avoiding the generation of that result altogether. Although this is also the focus of the work on factorized representations, their techniques work at the level of schemas and would not be able to
avoid generating the full result. To elaborate, consider the following query that does not do the projection:
\begin{lstlisting}[breaklines,basicstyle=\ttfamily]
   EdgesNP(ID1, PubID, ID2) :- AuthorPublication(ID1,PubID), AuthorPublication(ID2,PubID).
\end{lstlisting}
The query result for this query can be represented using the {\em f-tree} shown in Figure~\ref{fig:ftree} (T1), and the size of the factorization (F1) is linear in the size of the joining relations, in this case, the size of the \texttt{AuthorPublication} relation. This factorized representation is, in fact, equivalent to C-DUP; both of these use explicit nodes to represent the different {\tt PubID} elements. Another way to look at this is that, the query result here has a {\em multi-valued dependency}: \texttt{PubID} $\rightarrow\rightarrow $\texttt{ID1}, which these representations exploit.

However, both of these representations suffer from duplication since a pair of authors may share multiple \texttt{PubID}s. More specifically, although it is possible to generate the results of the second query with ``optimal delay tuple enumeration'' (Theorem 4.11~\cite{olteanu2015size}), the same pair of authors may be generated multiple times with different \texttt{PubID}s.
Projecting out the {\tt PubID} attribute results in the {\em f-tree} shown in Figure \ref{fig:ftree} (T2). This {\em f-tree} is, however, equivalent to doing the join, removing the duplicates, and generating the full result, since for every author, we must list out all of its co-authors (as shown in Figure~\ref{fig:ftree} (F2)).

\section{DEDUP-2 Greedy Algorithm}
\label{app:dedup2}

Conducting deduplication for outputting the DEDUP-2 representation turns out to be significantly more challenging than DEDUP-1 and we only present one algorithm for doing so (a few other variants
that we tried turned out to be too complex and inefficient, without any performance benefits).
Because the Virtual Nodes First algorithm reasons about a virtual node at a time, it turns out to be most amenable to adding such edges between
virtual nodes. 
Figure \ref{fig:dedup2_examples} shows an example of the different outcomes after incrementally introducing a new virtual node $V$ to a condensed graph containing only a single virtual node currently ($V_1$). Figure~\ref{fig:dedup2_examples}c shows the resulting graph after one applies our DEDUP-2 greedy algorithm to the condensed duplicated graph shown in Figure~\ref{fig:dedup2_examples}a. This graph now includes edges \textit{between} virtual nodes.
The addition of this new \textit{type} of edge to the mix immediately makes deduplication substantially more complex, as \textit{more invariants} need to now be checked in attempting to add the new virtual node into the deduplicated partial graph in order to maintain correctness.

Below we present an algorithm for adding a new virtual node $V$ into a partially constructed, deduplicated graph.
This algorithm does not add direct edges between real nodes; instead we introduce the notion of a \textit{singleton} virtual node both for the purpose of implicitly adding direct edges as well as to deal with correctness issues. A \textit{singleton} virtual node is one with only a single real node attached to it.

\topic{Step 1: Identifying violations} We first identify the set of virtual nodes that overlap with $V$ and may potentially lead to a violation. Let $\calV = \{V_1, ..., V_n\}$ denote all the virtual nodes in the partial condensed graph constructed so far, that share at least 1 real node with $V$; these are all the virtual nodes that we need to check for violations when adding $V$.
There are two types of violations that could potentially arise: (1) there exists a virtual node $V_1$ such that $|V \cap V_1| > 1$, (same as in all of the other representations and algorithms) as well as (2) there exists two \textit{other} virtual nodes $V_1$ and $V_2$ that are connected to \textit{at least one
    common virtual node} $C$ where $|V_1 \cap V_2| > 0$. More intuitively, violation (1) says there should be no overlap of more than 1 between any two virtual nodes, while (2) says that at any point in the graph, the virtual \textit{neighbors} of any one virtual node should have zero overlap \textit{with each other}.
    The reason (2) constitutes a violation comes up in iteration (\textit{getNeighbors()}) in this representation, where that scenario would lead to the same real node being returned as a neighbor multiple times.

\topic{Step 2: Edges between virtual nodes}
Let $V_1 \in \calV$ denote the virtual node with the highest overlap with $V$.

If $V$ has a high overlap with $V_1$, then removing this violation by adding direct edges (as above) could result in the addition of many direct edges between the real nodes (as seen in the example in Figure \ref{fig:dedup2_examples}b).
Hence, if $|V \cap V_1| \ge 1 $ we split both $V$ and $V_1$ and create 4 different virtual nodes so as to correctly incorporate $V$'s nodes into the partially deduplicated graph. These virtual nodes are: (1) $W_1 = V_1 \cap V$, (2) $W_2 = V_1 - W_1$, (3) $W_3 = V - W_1 - \cup N(V_1)$, (4) $W_4 = V - W_1 -
W_3$, where $\cup N(V_1)$ is the union of real nodes in $V_1$'s virtual neighborhood (some of these might be empty and would not be created).

To explain the intuition behind the above splits, we must explain how the algorithm works. The basic idea is to observe the current state of the deduplicated graph and see which virtual nodes need to be \textit{split} in order to make way for correctly adding $V$; this results into $W_1$ and $W_2$. After that we need to keep track of which edges need to be maintained, while recursively \textit{adding} in the portions of $V$ that could potentially cause violations ($W_3$ and $W_4$).
The most important part in the implementation of this algorithm is to not actually add \textit{any} edges between virtual nodes unless we are certain that these edges will not lead to any violations of the aforementioned invariants. To achieve this, we maintain a data structure \texttt{m} that includes all the edges that need to be added at any point in the execution, and after all the appropriate checks are made, only then are those edges physically added.

Processing each new virtual node $V$ can be described in smaller sub-steps:
\begin{enumerate}[noitemsep,nolistsep]
  \item \textbf{Substep 1}: Since $W_1 = V \cap V_1$, it will need to be a separate virtual node that both $V$ and $V_1$ will need to haves edges to. The intention of the first split is replacing $V_1$ with $W_1$ and $W_2$ where $W_1 \leftrightarrow W_2$. After this split, $W_1$ and $W_2$ also need to be connected to all the previous neighbors of $V_1$. This split can be applied immediately in the deduplicated graph as it does not alter the properties of the graph in any way. We keep track of the fact that $W_1$ and $V - W_1$ need to be connected after all checks are done by adding this potential edge in \texttt{m}.
  \item \textbf{Substep 2}: For simplicity, let virtual node $W_3' = V - W_1$ which includes the rest of the real nodes that are \textit{not} included in the initial split. We check if there are are any neighbors of $V_1$ that have \textit{any} overlap with $W_3'$, and if so, these nodes will constitute
  $W_4$, and the rest of the nodes will constitute $W_3$. If $W_4$ is not empty, we add the edge $W_3 \leftrightarrow W_4$ to \texttt{m} as well as $W_3 \leftrightarrow W_1$. If however there is a previous constraint in \texttt{m} for the virtual edge $W_3' \leftrightarrow  W_1$, then that also needs to be split into two constraints inside \texttt{m}.
  \item \textbf{Substep 3}: Recursively call the above on $V = W_3$ and then on $V = W_4$, passing the same \texttt{m} into every call.
  \item \textbf{Substep 4}: Physically add in all the edges that are described in \texttt{m} and clean up any virtual nodes that need to be deleted.
\end{enumerate}

\noindent
Please refer to Algorithm~\ref{vnodes2} in Appendix~\ref{app:pseudocodes} for the pseudo-code.

\topicul{Complexity} The runtime complexity of this algorithm is hard to calculate precisely, but is upper bounded by $O(n_r*d^8)$.

\section{Details on Experimental Setup}
In this section we describe the experimental setup in terms of the way the datasets that we used were generated. We also include details on the database schema used for some of the real datasets which we experimented on, as well as provide examples of the query extraction SQL generated by our system which efficiently extracted the C-DUP representation of these graphs.

\subsection{Generation of Small Synthetic Datasets}
\label{app:smalldatasets}
We briefly describe our algorithm for generating small synthetic datasets for the detailed experiments. We needed
the ability to generate a series of synthetic graphs so that we can better understand the differences between
the representations and algorithms on a wide range of possible datasets, with varying numbers of real nodes and virtual nodes, and varying degree distributions and densities.
However, we could not use any of the existing random graph generators for this purpose; this is because we need the graphs in a condensed representation.
Instead, we built a synthetic graph generator based on
the Barab{\`a}si--Albert model~\cite{albert2002statistical} (also called the {\em preferential attachment model}).
that takes as input: the number of real nodes ($n_1$), the number of \textit{virtual nodes} ($n_2$), as well as the mean $m$ and the standard deviation $sd$ that define the normal distribution from which we draw the random sizes (degrees) of the virtual nodes.

We sketch the algorithm here: 
\begin{enumerate}[noitemsep,nolistsep]
  \item Add all real nodes into the graph at once and \textit{generate} all virtual nodes and their sizes by sampling the ($m,sd$) normal distribution.
  \item \textbf{Initial Splits:} For every virtual node $vn$, \textit{split} $vn$ into $s_1,s_2$ with probability relative to its size.
  \item \textbf{Initial Batch Random Assignment:} Add $15\%$ of the virtual nodes to the graph, and attach real nodes to them at random.
  \item \textbf{Random or Preferential Attachment:} For each $v_i$  with size $x$ in the remaining virtual nodes, if $v_i$ was derived from a \textit{split}, then with probability $35\%$, randomly assign real nodes to $v_i$. Otherwise, 
    randomly select a real node $r$ in the graph that currently has a degree of $d(r) \ge x$. Let $s$ the set of $r$'s neighbors. Assign probabilities to each neighbor $s_i$ as such: $P_i = d(s_i)^2/ \sum{d(s_i)^2}$. Until $d(r) = x$, remove a real node $s_i$ from $s$ with probability \textit{counter-proportional} to $P_i$. Real nodes with a high degree, and therefore high $P_i$ value, are more likely to remain in $s$ and thus be attached to $v_i$.
  \item \textbf{Cleanup:} Merge the virtual nodes that derived from splitting in step $3$,  back into their one original virtual node.
\end{enumerate}
This algorithm can be used to generate a graph with similar degree distributions as those generated by the commonly-used {\em preferential attachment model}~\cite{albert2002statistical}, while also preserving the local densities typically seen in real-world networks (which the naive preferential
attachment model does not preserve~\cite{leskovec2007graph}).

\begin{figure}[t]
\begin{center}
\includegraphics[width=0.45\textwidth]{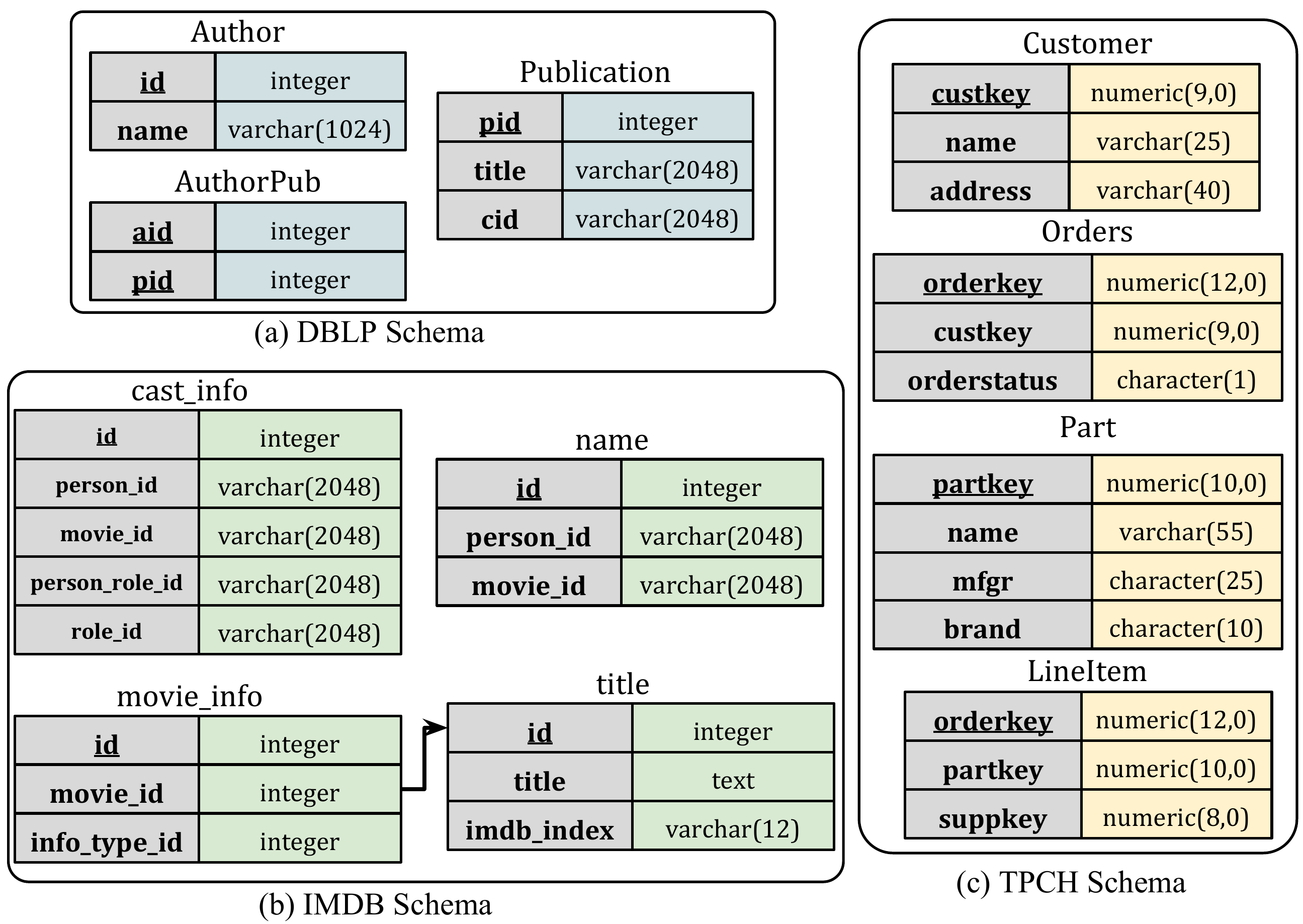}
\caption{Database Schemas: If not explicitly shown, foreign key constraints for each attribute (if any) refer to the the primary key attribute in a different table with the same name.}
\label{fig:schemata}
\end{center}
\end{figure}

\begin{figure}[t]
\begin{center}
\includegraphics[width=0.45\textwidth]{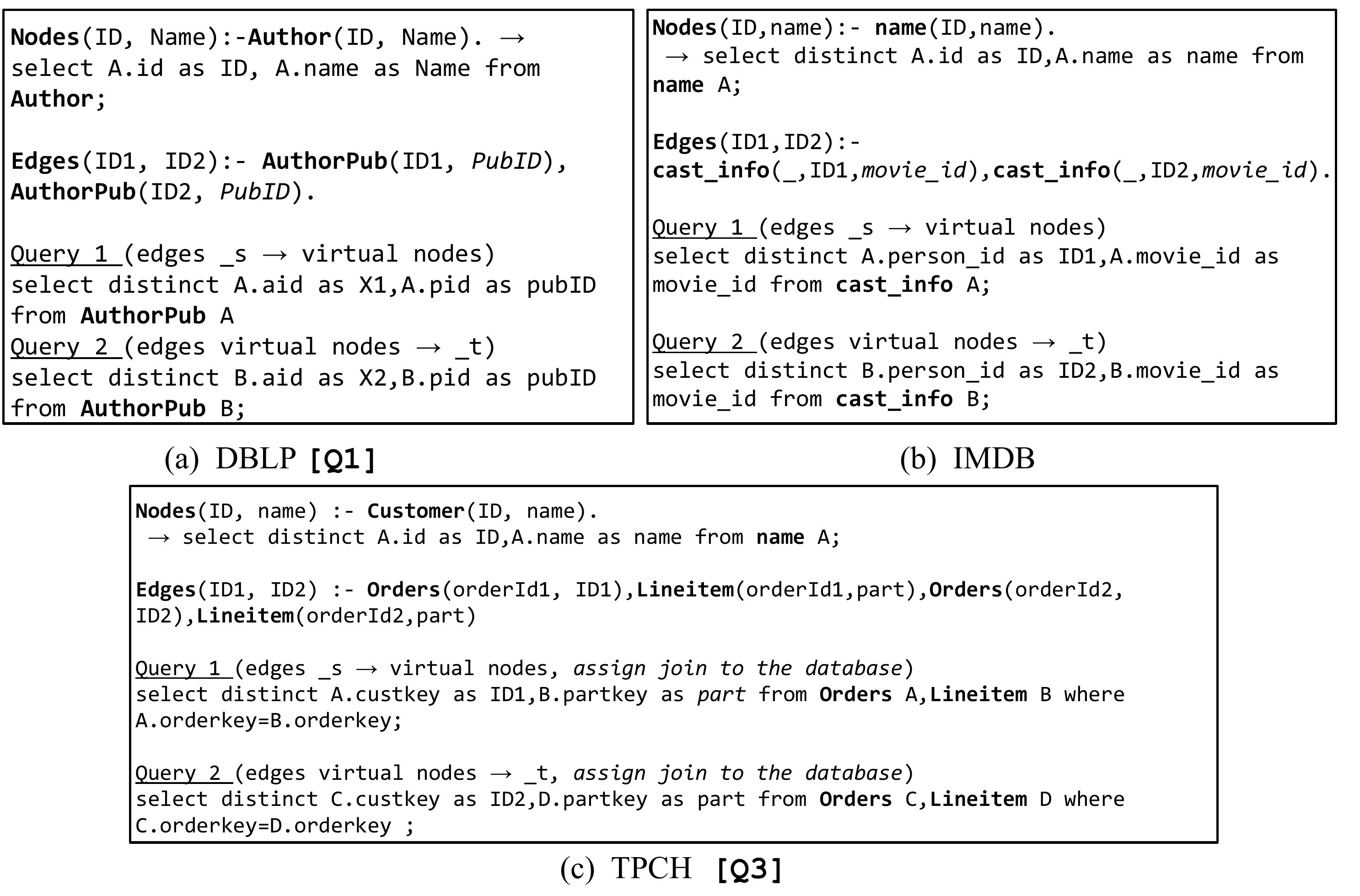}
\caption{The SQL generated from the system for a few of the graphs we used in our experiments}
\label{fig:sqlQueries}
\end{center}
\end{figure}

\subsection{Generation of Large Datasets}
\label{app:largedatasets}
The Layered\_1, Layered\_2, Single\_1, Single\_2, datasets are  synthetically generated multi-layer and single-layer datasets. Both Layered\_1 and
Layered\_2 have the same layer structure as the TPCH example, as shown in Figure~\ref{fig:query_3-4}. The way these were generated was by generating
database tables while adjusting the cardinality of the join condition attributes for those tables.The tables were created by randomly generating
values in a range of integers (uniformly distributed). More information about the generated datasets can be seen in Table~\ref{tab:selectivities}.
The numbers in column joinSelectivities show the selectivity of each join that would be required for creating the full graph. In Layered\_1 for
instance (since the join structure is the same as TPCH), there are 3 joins that are required across 2 generated tables. Let those tables be $A,B$; the joins required here were $A \Join B$ which had selectivity $0.05$, $B \Join B$ with a selectivity of $0.1$ and again $B \Join A$ with selectivity
$0.05$. The definition of \textit{selectivity} that we use here for a particular join on an attribute $a$ of a table $A$ is $selectivity = distinct\_a/|A|$, where $distinct\_a$ the distinct number of unique values of $a$.

\begin{table}[t]
\center
\csvreader[tabular=|c|c|c|c|,
    table head=\hline \textbf{Dataset} & \textbf{Nodes} & \textbf{Edges} & \textbf{Join Selectivities} \\\hline,
    late after line=\\\hline]%
{numbers/selectivities.csv}{dataset=\dataset,nodes=\nodes,edges=\edges,joinSelectivities=\selectiv}%
{\dataset & \num[group-separator={,}]{\nodes} & \num[group-separator={,}]{\edges} & \selectiv}
\caption{Selectivities of synthetically generated multi-layer and single layer datasets. The nodes and edges sizes shown here are of the C-DUP representation of these graphs.}
\label{tab:selectivities}
\end{table}

\subsection{Database Schemas and Generated SQL}
We experimented on various real world datasets, some of which include DBLP, IMDB, and TPCH. The DBLP database  includes authors and their publications to conferences, the IMDB database includes information about movies, the actors that acted in them as well as directors, crew etc. The TPCH database includes information about Customers and Orders they have placed, as well as information about said orders, suppliers etc.  Figure~\ref{fig:schemata} shows subsets of the schemas for these databases. It's important to note that looking at these schemas, users can intuitively come up with various types of graphs that could be extracted and analyzed from these datasets; e.g. looking at the TPCH Schema (Figure~\ref{fig:schemata}c), one can see that LineItem has an attribute for the supplier (\texttt{LineItem.suppkey}), and therefore, since there also exists a Supplier relation, a graph of suppliers could also be extracted. For the purposes of our experiments, we have extracted the graphs described in Section~\ref{sec:experiments}
(the IMDB and DBLP graphs in Section~\ref{sub:smalldatasets} and the TPCH graph in Section~\ref{sec:large} ).The extraction queries in our Datalog DSL and the resulting SQL queries for these graphs can be seen in Figure~\ref{fig:sqlQueries}

\section{Algorithm Pseudocodes}
\label{app:pseudocodes}
Here, we sketch the pseudo-codes for some of the different algorithms presented in the main body of the paper.

\makeatletter
\def\BState{\State\hskip-\ALG@thistlm}
\makeatother

\begin{algorithm}
\fontsize{8pt}{8pt}\selectfont
\caption{BITMAP-2}\label{bmp2}
\begin{algorithmic}[1]
\Procedure{BMP2}{$graph$,$ordering$}
\State $srted \gets graph.vertices$.sortByDuplication($ordering$)
\State $seen \gets hashSet()$
    \For{each real node $rn$ in $srted$}
        \State $virtSet \gets $ greedySetCover($rn$)
        \For{each virtual node $v \in virtSet$}
                \For{each $index$, real node $rn_2$ in $v$.getOutNeighbors()}
                    \If{$rn_2 \notin seen$}
                        \State $v$.getBitmap($rn$).setBitAt($index$)
                        \State $seen$.add($rn_2$)
                    \Else
                        \State $chosen \gets false$
                        \For {each bitmap $bmp$ in $v$.getBitmaps()}
                            \If {$bmp$.getBitFor($rn$) $== 1$}
                            \State $chosen \gets true$
                            \State \textbf{break}
                            \EndIf

                        \EndFor
                        \If{$!chosen$}
                          \State $v$.removeBitMapFor($rn$)
                          \State removeEdge($rn$,$v$)
                        \EndIf
                    \EndIf

                \EndFor
                \State $v$.rebuildBitmapIndex()
        \EndFor
        \State $seen$.clear()
    \EndFor
\EndProcedure
\end{algorithmic}
\end{algorithm}

\newpage


\makeatletter
\def\BState{\State\hskip-\ALG@thistlm}
\makeatother

\begin{algorithm}[h]
  \fontsize{7.5pt}{7.5pt}\selectfont
\caption{BITMAP-1}\label{bmp1}
\begin{algorithmic}[1]
\Procedure{BMP1}{$graph$}
\State $seen \gets hashSet()$
    \For{each real node \emph{rn} in $graph.vertices()$}
        \For{each virtual node \emph{vn} in $rn.getOutNeighbors()$}
            \For{each $index$, real node $rn_2$ in $vn.getOutNeighbors()$}
                \If{$rn2 \notin seen$}
                    \State set bit at $index$ of bitmap $vn$.bitMaps.get($rn$)
                    \State $seen$.add($rn2$)
                \EndIf
            \EndFor
        \EndFor
        \State $seen$.clear()
    \EndFor

\EndProcedure
\end{algorithmic}
\end{algorithm}



\makeatletter
\def\BState{\State\hskip-\ALG@thistlm}
\makeatother

\begin{algorithm}
  \fontsize{8pt}{8pt}\selectfont
\caption{Greedy Virtual Nodes First I}\label{vnodes1}
\begin{algorithmic}[1]
\Procedure{virtNodesFirst1}{$graph$,$ordering$}
\State $processed \gets hashSet()$
\State $srtd \gets $ orderVirtualNodes($ordering$)
\For{each virtual node $v \in srtd$}
    \State $relevant \gets $ getRelevantVNodes($v$)
    \State $moreDedup \gets $\textbf{true}
    \While{$moreDedup$}
        \State $moreDedup \gets $\textbf{false}
        \State $intersections \gets $ getIntersections($v$,$relevant$)

        \For{each $s \in relevant$}
            \State $C_i \gets intersections$.get($s$)
            \If{$|C_i|$ $> 1$}
                \State $moreDedup \gets $\textbf{true}
                \For{each real node $rn \in C_i$}
                    \State $R,V,DirectEdges\gets $ maxBenefitRatio($rn$)
                \EndFor
            \EndIf
        \EndFor
        \If{$R \neq $\texttt{Null}}
            \State $graph$.removeEdge($R,V$)
            \State addDirectEdges($R,V,DirectEdges$)
        \EndIf
    \State $processed$.add($v$)
    \EndWhile
\EndFor
\EndProcedure
\end{algorithmic}
\end{algorithm}


\makeatletter
\def\BState{\State\hskip-\ALG@thistlm}
\makeatother

\begin{algorithm}
  \fontsize{6pt}{6pt}\selectfont
\caption{Greedy Real Nodes First}\label{realNodes}
\begin{algorithmic}[1]
\Procedure{realNodesFirst}{$graph$,$ordering$}
\State $V' \gets hashSet()$
\State $V'' \gets hashSet()$

\State $X \gets hashSet()$
\State $srtd \gets$ graph.sortRealNodesByDuplication($ordering$)
\For {each real node $rn \in srtd$}
    \State initialize($V''$,$rn$)
    \While {$V'' \neq \varnothing$ }
        \State $maxBenefitVNode \gets $ getMaxBenefitCostRatioVNode()
        \If{$maxBenefitVNode \neq  $\texttt{Null}}
            \State $V \gets hashSet()$
            \For{real node $rn_2 \in maxBenefitVNode$.getOutNeighbors()}
                \State $V$.add($rn_2$)
                \State $graph$.removeEdge($rn$,$rn_2$)
            \EndFor
            \State $V'$.add($maxBenefitVNode$)
            \State $V''$.remove($maxBenefitVNode$)
            \State $VCapX \gets V \cap X$

            \For{each pair $a,b$ of $a \in V-VCapX$ and $b$ in $VCapX$}
                \If{$!$existsEdge($a,b$)}
                    \State $graph$.addEdge($a,b$)
                \EndIf
            \EndFor

            \For{each real node $s \in V$}
                \If{$s$.equals($rn$)}
                    \State{$X$.add($s$)}
                \EndIf
            \EndFor

            \For{each real node $s$ in $VCapX$}
                \State $graph$.removeEdge($s$,$maxBVNode$)
            \EndFor
        \Else
            \For{real node $vn$ in $V''$}
                \State removeEdge($rn$,$vn$)
            \EndFor
        \EndIf
    \EndWhile
\State $X$.clear()
\EndFor
\EndProcedure
\end{algorithmic}
\end{algorithm}


\makeatletter
\def\BState{\State\hskip-\ALG@thistlm}
\makeatother

\begin{algorithm}
  \fontsize{8pt}{8pt}\selectfont
\caption{Greedy Virtual Nodes First II}\label{vnodes2}
\begin{algorithmic}[1]
\Procedure{VirtNodesFirst2}{$graph$,$ordering$}
    \State $srtd \gets $orderVNodes($ordering$)
    \For{each virtual node $vn$ in $srtd$}
        \State $constraints \gets hashMap()$
        \State ResolveVirtualNode($v$,$constraints$)
    \EndFor
\EndProcedure

\Procedure{ResolveVirtualNode}{$v$,$constraints$}
    \State $relevant \gets $getRelevantVNodes($v$)
    \State $HV \gets $highestOverlap($v,relevant$)

    \State $W_1 \gets $intersect($HV,v$)
    \State $w_1 \gets $createVirtNode($W_1$)
    \If{$W_1 \neq \varnothing $}
        \State $W_2 \gets HV-W_1$
        \If{$W_2 \neq \varnothing $}
            \State $w_2 \gets $createVirtNode($W_2$)
            \State addVirtualEdge($w1,w2$)
        \EndIf

        \For{each virtual node $vn \in HV.$getVirtualNeighbors()}
            \State addVirtualEdge($w1,vn$)
            \State addVirtualEdge($w2,vn$)
        \EndFor

        \State $W_3' \gets v - W_1$

        \State $W_3 \gets W3' - $NUnion($HV$)
        \If{$W_3 \neq \varnothing $}
            \State $w_3 \gets $createVirtNode($W_3$)
            \State addConstraint($constraints,w_3,w_1$)
        \EndIf

        \State $W_4 \gets W3' - W3$
        \If{$W_4 \neq \varnothing $}
            \State $w_4 \gets $createVirtNode($W_4$)
            \If{$W_3 \neq \varnothing $}
                \State addConstraint($constraints,w_4,w_3$)
            \EndIf
        \EndIf

        \If{$v$ exists in $constraints$}
            \State splitConstraintsFor($constraints,v$)
        \EndIf

        \If{$w_4 \neq $\texttt{Null}}
            \State \textbf{ResolveVirtualNode}($w_4,constraints$)
        \EndIf
        \If{$w_3 \neq $\texttt{Null}}
            \State \textbf{ResolveVirtualNode}($w_3,constraints$)
        \EndIf
    \EndIf
    \State $graph$.removeVirtualNode($v$)
    \If{initialCall}
        \State $graph$.addVirtualedgesIn($constraints$)
    \EndIf
\EndProcedure
\end{algorithmic}
\end{algorithm}

\balance

\end{document}